\documentclass[aps,prd,onecolumn,superscriptaddress,showpacs,floatfix,preprintnumbers,nofootinbib,10pt]{revtex4-2}

\usepackage[dvipsnames]{xcolor}      
\definecolor{lcolor}{rgb}{0.5,0,0}
\definecolor{citcolor}{rgb}{0,0.0,1}
\usepackage[breaklinks,colorlinks,urlcolor=blue,citecolor=citcolor,linkcolor=lcolor,linktoc=all]{hyperref}
\usepackage{xcolor}
\usepackage{graphicx}	
\graphicspath{{./figures/}}
\usepackage[normalem]{ulem}
\usepackage{amsmath} 
\usepackage{amssymb}
\usepackage[ragged]{footmisc}
\usepackage{slashed}
\usepackage{dsfont}

\usepackage{tikz}
\usepackage[customcolors]{hf-tikz}
\usepackage[compat=1.1.0]{tikz-feynman}
\usepackage{silence}
\newcommand{\eq}{Eq.~}
\newcommand{\ud}{\mathrm{d}}
\newcommand{\as}{\alpha_\mathrm{s}}

\newcommand{\WFqgg}[1]{\psi^{q\rightarrow qgg}_{#1}}

\newcommand{\measure}[1]{\widetilde{\mathrm{d}#1}}

\begin{document}

\title{Two-Loop DGLAP Splitting Functions from Light Cone Perturbation Theory}

\preprint{HIP-2026-2/TH}

\author{Tuomas Lappi}
\address{
Department of Physics, 
P.O. Box 25, FI-40014 University of Jyväskylä, Finland
}
\address{
 Helsinki Institute of Physics,
P.O. Box 64, FI-00014 University of Helsinki, Finland
}
\author{Risto Paatelainen}
\address{
Department of Physics 
P.O. Box 64, FI-00014 University of Helsinki, Finland
}
\address{
 Helsinki Institute of Physics,
P.O. Box 64, FI-00014 University of Helsinki, Finland
}
\author{Mikko Seppälä}
\address{
Department of Physics 
P.O. Box 64, FI-00014 University of Helsinki, Finland
}
\address{
 Helsinki Institute of Physics,
P.O. Box 64, FI-00014 University of Helsinki, Finland
}

\begin{abstract}
    We perform a two-loop calculation in Light Cone Perturbation Theory (LCPT) to evaluate the next-to-leading order nonsinglet splitting function. Our calculation demonstrates the methodology and feasibility of performing higher order calculations in LCPT. Since in Hamiltonian perturbation theory the longitudinal $k^+$ momentum is always positive, poles in $1/k^+$ can be regularized by a simple cutoff which cancels in physical results, without any associated ambiguities. For transverse momentum integrals we use dimensional regularization. Developing methods for loop calculations in LCPT paves the way for a systematical, automatizable procedure for precision calculations in this framework with a transparent physical partonic interpretation. This can provide  a standard framework in higher order calculations in the gluon saturation regime of QCD.
\end{abstract}

\maketitle

\section{Introduction}
Perturbative Quantum Chromodynamics (QCD) describes the structure of strongly interacting bound states in terms of their quark and gluon constituents, \emph{partons}. However,  relating experimental observables to the fundamental theory is complicated by the fact that partons are never observed as such in the laboratory. In stead, the predictive power of the theory is achieved by a factorization of universal nonperturbative aspects of the bound states and perturbatively calculable partonic interactions. 

Such a factorization between perturbative and nonperturbative physics, in a general sense, can take many forms. The most common of these is collinear factorization, where the nonperturbative description is provided by parton distribution functions (PDFs). Although PDFs are nonperturbative, they are universal, and importantly their dependence on the momentum scale probed in the scattering process can be calculated perturbatively using the DGLAP~\cite{Gribov:1972ri,Dokshitzer:1977sg,Altarelli:1977zs} equations. For more differential or exclusive measurements the concept of PDFs can be generalized e.g. to transverse momentum dependent distributions (TMDs), generalized parton distributions (GPDs) et cetera~\cite{Lorce:2025aqp}. 

In the regime of high scattering energies (i.e. at small values of the target momentum fraction $x$), especially for large nuclei, one enters in a regime of gluon saturation, where  the number density of gluonic states in the target is so high that a perturbative description of it in terms of PDFs is not adequate. Here, the small-$x$ gluons in the high energy nucleus can be described in terms of a classical color field, in the Color Glass Condensate (CGC) effective theory~\cite{Gelis:2010nm}.  However, even then, the language is fundamentally that of partons. Experimentally the CGC field can be probed by a scattering  the dense nucleus with a dilute probe,  e.g. a large-$x$ parton in a forward rapidity hadronic cross section, or a (potentially virtual) photon in deep inelastic scattering or an ultraperipheral heavy ion collision. In these cases the calculation of the cross section requires a partonic description of the probe. Such dilute-dense calculations have indeed seen  remarkable advances in recent years, with a growing number of calculations performed at next-to-leading order (NLO) in a perturbative description of the probe (see e.g.~\cite{Chirilli:2011km,Boussarie:2016bkq,Beuf:2016wdz,Beuf:2017bpd, Hanninen:2017ddy,Beuf:2021srj,Mantysaari:2021ryb, Taels:2022tza,Beuf:2024msh}). Derivations  of the associated B-JIMWLK high energy evolution equations~\cite{Balitsky:1995ub,Kovchegov:1999yj,Kovchegov:1999ua} at NLO or higher accuracy in fact proceed by a calculation of an even  higher order scattering process in a soft gluon emission limit~\cite{Balitsky:2007feb, Balitsky:2013fea,Kovner:2013ona,Caron-Huot:2016tzz}. 

Partonic calculations at higher orders are thus needed. This, however, forces us to face a technical issue. The most economical way to perform a loop calculation in QCD is to do so in a covariant, such as Feynman, gauge where one can take advantage of Lorentz symmetry in simplifying loop integrals. Partons, however, are not really fundamentally  covariant gauge objects. They are physical states probed in a collision process at equal light cone time, with only physical polarizations for gluons. Thus the physical interpretation in terms of partons requires one to fix the axial gauge $A^+=0$ to remove unphysical polarizations of gluons, and get rid of longitudinal Wilson lines required to make two-body operators (parton densities) gauge invariant. In particular CGC calculations of dilute-dense scattering processes are practically always performed in axial  gauge; some covariant gauge calculations exist at leading order~\cite{Kovchegov:1997ke,Blaizot:2004wu,Blaizot:2004wv,Gelis:2005pt} but would be prohibitively complicated at higher orders. 

The advantage of axial (light cone) gauge has of course been realized for a long time, and was at a point actively studied (e.g.~\cite{Pritchard:1978ts,Leibbrandt:1983zd, Leibbrandt:1999ag}). In particular a crucial reference for our calculation here is the DGLAP splitting function calculation of  Curci, Furmanski and Petronzio~\cite{Curci:1980uw}. Covariant (Feynman) perturbation theory calculations at higher orders have, however, gradually gravitated towards covariant gauge. The undesirability of axial gauge in this context is often attributed to the spurious $1/k^+$ pole in axial gauge~\cite{Belitsky:2002sm}, which is claimed to present  difficulties in regularization or be otherwise  ambiguous. More recent work points, however, towards a solution to this issue in the form of Hamiltonian perturbation theory. In ``old-fashioned'' perturbation theory in light cone gauge, i.e. light cone perturbation theory~\cite{Kogut:1969xa,Bjorken:1970ah,Brodsky:1997de}, one does not integrate over this pole, because the momentum $k^+$ is strictly positive. The $1/k^+$ divergence can then be regularized, e.g. with a cutoff (see~\cite{Altinoluk:2025tms} for a different regularization approach), and unphysical effects resulting from this pole will cancel in physical observables. This is the approach followed with success in  recent NLO CGC calculations mentioned above. 

This discussion leads us to the main goal of our present work. We will demonstrate with an explicit calculation that two-loop calculations in LCPT are possible, feasible and free of  ambiguities that they may have sometimes been conjectured to possess. As an example we rederive here the NLO non-singlet (valence quark) DGLAP splitting function. We regularize transverse momentum integrals by dimensional regularization, and introduce for the $k^+$ integrals a cutoff which cancels in the final result. We see this as just a first step towards a more systematical approach to higher order calculations in LCPT, including a thorough automatization using modern computer algebra tools. We intend to return to such developments in future work.
We will first discuss the definition of a parton distribution function and calculate the leading order quark-to-quark splitting function in Sec.~\ref{section: LO derivation}. Then in Sec.~\ref{sec:nlo} we list the diagrams that are needed at NLO and discuss general features of the calculation. We then  delve into more details and arrive at the final result in Sec.~\ref{sec:details}, before concluding and outlining potential further developments in Sec~\ref{sec:conc}.

\section{Parton distributions and DGLAP evolution in LCPT}\label{section: LO derivation}

Light cone wave functions (LCWF's) describe the decomposition of asymptotic, dressed partons into bare parton states that are probed in hard, short time scale scattering processes. In the jargon of the field, one often talks about ``splitting'' wavefunctions. It might feel  intuitively clear that such splitting wavefunctions should be related to the DGLAP splitting functions. This is particularly clear and explicit at one loop level (see e.g. the forms in Ref~\cite{Lappi:2016oup}), where  the splitting function is given by the square of the splitting wavefunction. For proper quantitative work at higher loop orders, one must however define things more explicitly. 

Often, in the more pedagogical and approachable parts of the literature, one approaches the issue by starting with a physical process such as Deep Inelastic Scattering (DIS). When  calculating the cross section at NLO  with massless on-shell incoming particles one then encounters a collinear singularity, which is then absorbed into a DGLAP evolution of the PDF's. This captures well the correct physics in the sense that the purpose of DGLAP evolution is to resum large collinear logarithmic corrections to cross sections. However, this approach  is not manifestly universal and process independent, and there are many reasons to prefer a renormalization approach based on an operator definition of the PDF (as argued rather forcefully e.g. in Ref.~\cite{Collins:2011zzd}).
We will here use the operator-based definition of the PDF. But our purpose is not to dwell on the fine points of renormalization too much here, but to find a direct route to extract the splitting function from the LCWF. Thus we will be very brief here, and present a slightly more precise discussion of the renormalization of parton distributions in Appendix~\ref{app:renormalization}.

Our starting point is to define the light cone gauge parton distribution function of a parton of type $i$ in a hadron $h$ 
%at a regularization  scale $\bar{\mu}$ 
as the expectation value of the bare parton number operator in the hadronic state $|h\rangle$: 
\begin{equation}\label{PDF definition}
    f^0_{i/h}(\mu,x)\equiv\frac{1}{4\pi x}\int\frac{\mathrm{d}^{D-2}\mathbf{k}}{(2\pi)^{D-2}}\frac{\left\langle h(p^+)\right|a_i^{\dagger}(xp^+,\mathbf{k})a_{i}(xp^+,\mathbf{k})\left|h(p^+)\right\rangle}{\left\langle h(p^+)|h(p^+)\right\rangle}.
\end{equation}
Here by a ``bare'' parton, measured by the bare number operator $a_i^{\dagger}(xp^+,\mathbf{k})a_{i}(xp^+,\mathbf{k})$, state we mean a non-interacting Fock state in comparison with the full interacting theory hadronic state. 
Bare partons are not physical eigenstates of the theory, and a bare parton distribution is not a physically measurable quantity. Thus, it is not surprising that the bare PDF defined by Eq.~\eqref{PDF definition} is ultraviolet (UV) divergent, and to make sense of it we have had to regularize it using dimensional regularization in $D=4-2\varepsilon$ dimensions. In order for the action to be dimensionless, we then also have to accompany every power of the coupling by a dimensionful scale, so that the expansion parameter in a perturbative expansion is $\bar{\mu}^\varepsilon g$, with  $\bar{\mu}=4\pi e^{-\gamma_E}\mu$ and the bare coupling $g$ dimensionless. This makes our bare PDF depend, starting from one loop, on the regularization scale $\mu$. The physical PDF needs to be renormalized by subtracting the UV divergence, which removes the singularities in the $D\to0$ limit. This will   replace the  dependence on the regularization scale $\mu$ by a dependence on the renormalization or factorization scale $\mu_F$.

Since logarithms of the renormalization scale always accompany $1/\varepsilon$ poles, one can read off the splitting functions from the coefficients of these poles, in a way that we will make more explicit first through a one-loop example, and then with a full definition at two loops in  Eq.~\eqref{eq: pdf in terms of splitting functions}.
It might seem counterintuitive that the splitting function, which is after all used to resum collinear divergences, is related to an UV divergence. One way to interpret this is an effective theory approach. The PDFs describe the soft sector of the calculation, and are defined with an UV cutoff. Often the integrated PDF is indeed defined as the integral of an unintegrated distribution up to an explicit cutoff~\cite{Kovchegov:2012mbw}, although this is not exactly the approach we will follow here. The hard scattering cross sections, on the other hand, are treated in an UV effective theory, where partons  are treated as on-shell, leading to collinear divergences. 
In reality physical partons are always off shell since they are constituents of a bound state, and cross sections involving hadrons are not collinearly divergent, but merely feature large logarithms of hard scales divided by some  soft hadronic scale. 
The UV divergence of the soft sector and the collinear divergence of the hard sector are just artefacts of kinematical approximations appropriate for the soft and hard regimes, and must therefore cancel each other in the full theory. This connects the UV divergence of the operator definition of the PDF to collinear large logarithms in cross sections. We will first see how the renormalization of the UV divergence works in practice at one loop.

% UV and collinear divergences are renormalized at the same scale, $\mu_F$, using the MS-bar scheme definition
% \begin{equation}
%    \bar{\mu}_F=4\pi e^{-\gamma_E}\mu_F,
%\end{equation}
%with regularization carried out with dimensional regularization in $D=4-2\varepsilon$ %dimensions.

In the definition of the PDF, Eq.~\eqref{PDF definition}, the hadron carries longitudinal momentum $p^+$, of which the measured parton carries a fraction $x$. The operator $a_i^{\dagger}(\mathbf{k},xp^+)$ is the light cone creation operator that creates a bare particle of type $i$ with  momentum $\vec{k}=(xp^+,\mathbf{k})$. The operator $a_i$ is the corresponding annihilation operator. They satisfy the canonical commutation relation
\begin{equation}
    [a_i(\vec{k}),a_i^{\dagger}(\vec{p})]_\pm=2k^+(2\pi)^{D-1}\delta^{(D-1)}(\vec{k}-\vec{p}),
\end{equation}
where $[\dots]_\pm$ represents commutation or anti-commutation relations depending on particle $i$'s statistics. The normalization factor in Eq.~\eqref{PDF definition} is chosen in accordance with the normalization of the commutation relations so that integrating $f_{i/h}(\mu_F,x)$ over $x$ gives the total number of particles with longitudinal momentum $xp^+$. Note that we have left the colors and helicities of the partons implicit, and assume that helicities are summed over.

The hadronic state $|h(p^+)\rangle$ is, in general, a linear combination of dressed $n$-particle states weighted by some non-perturbative wave functions $\psi_n$ encapsulating the hadronic long distance physics:
\begin{equation}\label{eq: hadronic state}  |h\rangle=\sum_n\int\Big[\prod_{j=1}^n\widetilde{\mathrm{d}k}_j\Big]\psi_n(\{k_j\})|n\rangle_\mathrm{int}.
\end{equation}
However, since the quantities of interest that are perturbatively attainable, e.g. splitting functions, are independent of the hadronic state, we are free to extract out of Eq.~\eqref{eq: hadronic state} the contribution of a one-particle quark or gluon state and ignore the remaining parts. In this paper, we choose as the hadronic state a single interacting quark:
\begin{equation}
    |h(p^+)\rangle = |q(p^+)\rangle_\mathrm{int}.
\end{equation}
In LCPT we express our physical state as an expansion in bare states, which in the specific case of a single quark reads 
\begin{equation}\label{eq: expansion for a quark}
\begin{aligned}
    |q(\vec{p})\rangle_\mathrm{int}&=\sqrt{Z_q(p^+)}\Bigg[|q(\vec{p})\rangle+\int \measure{k}\measure{l}(2\pi)^{D-1}\delta^{(D-1)}(\vec{p}-\vec{k}-\vec{l}) \psi^{q\rightarrow qg}(\vec{p},\vec{k},\vec{l})|q(\vec{k})g(\vec{l})\rangle\\
    &+\frac{1}{\sqrt{2}}\int\measure{k}\measure{l}\measure{l'}(2\pi)^{D-1}\delta^{(D-1)}(\vec{p}-\vec{k}-\vec{l}-\vec{l}\,')\Big\{\psi^{q\rightarrow qgg}(\vec{p},\vec{k},\vec{l},\vec{l}\,')|q(\vec{k})g(\vec{l})g(\vec{l}\,')\rangle\\
    &+\psi^{q\rightarrow qq\bar{q}}(\vec{p},\vec{k},\vec{l},\vec{l}\,')|q(\vec{k})q(\vec{l})\bar{q}(\vec{l}\,')\rangle\Big\}+\dots\Bigg],
\end{aligned}
\end{equation}
where $Z_q$ is the quark state renormalization constant and $\psi^{q\rightarrow n}$ are the LCWFs corresponding to the different Fock states $|n\rangle$. The integration measure is defined as
\begin{equation}
    \int \measure{k}\equiv\int \frac{\mathrm{d}^{D-1}\vec{k}}{(2\pi)^{D-1}}\frac{\theta(k^+)}{2k^+}=\int_0^\infty\frac{\mathrm{d}k^+}{4\pi k^+}\int\frac{\mathrm{d}^{D-2}\mathbf{k}}{(2\pi)^{D-2}}.
\end{equation}

Combining the discussion thus far, we can write down, as an example, the distribution of bare quarks with momentum fraction $x$ in a dressed quark as
\begin{equation}\label{eq: PDF definition quark-quark}
\begin{aligned}
    f^0_{q/q}(\mu,x)&=\frac{1}{4\pi x} \int\frac{\mathrm{d}^{D-2}\mathbf{k}}{(2\pi)^{D-2}}\frac{{}_\mathrm{int}\langle q(p^+)|b^\dagger(xp^+,\mathbf{k})b(xp^+,\mathbf{k})|q(p^+)\rangle_\mathrm{int}}{{}_\mathrm{int}\langle q(p^+)|q(p^+)\rangle_\mathrm{int}},\\
\end{aligned}
\end{equation}
where $b^\dagger b$ is the quark number operator. 
Let us see in detail how the definition works out for the first terms of the expansion: inserting the Fock expansion Eq.~\eqref{eq: expansion for a quark} into the definition of the  PDF Eq.~\eqref{eq: PDF definition quark-quark} yields 
\begin{equation}\label{eq: quark in quark PDF expansion}
\begin{aligned}
    &f^0_{q/q}(\mu,x)=\frac{1}{4\pi x} \int\frac{\mathrm{d}^{D-2}\mathbf{k}}{(2\pi)^{D-2}}\frac{Z_q(p^+)}{{}_\text{int}\langle q(p^+)|q(p^+)\rangle_\text{int}}\Bigg[\langle q(\vec{p})|b^\dagger(\vec{k})b(\vec{k})|q(\vec{p})\rangle+ \int\measure{k'}\measure{k''}\measure{l}\measure{l'}(2\pi)^{D-1}\delta^{(D-1)}(\vec{p}-\vec{k}\,'-\vec{l})\\
    &\times(2\pi)^{D-1}\delta^{(D-1)}(\vec{p}-\vec{k}\,''-\vec{l}\,') \psi^{q\rightarrow qg}(\vec{p},\vec{k}\,',\vec{l})\left(\psi^{q\rightarrow qg}(\vec{p},\vec{k}\,'',\vec{l}\,')\right)^\dagger \langle q(\vec{k}\,'')g(\vec{l}\,')|b^\dagger(\vec{k})b(\vec{k})|q(\vec{k}\,')g(\vec{l})\rangle+\dots\Bigg]\\
    &=Z_q(p^+)\left[\delta(1-x)+\frac{1}{16\pi(p^+)^2x(1-x)}\int \frac{\mathrm{d}^{D-2}\mathbf{k}}{(2\pi)^{D-2}}\left|\psi^{q\rightarrow qg}(\vec{p},\vec{k},\vec{p}-\vec{k})\right|^2 +\dots\right],
\end{aligned}
\end{equation}
where we inserted the normalization and matrix elements:
\begin{equation}
\begin{aligned}
{}_\text{int}\langle q(\vec{p})|q(\vec{k})\rangle_\text{int}&=\langle q(\vec{p})|q(\vec{k})\rangle=2p^+(2\pi)^{D-1}\delta^{(D-1)}(\vec{p}-\vec{k}),\\
    \langle q(\vec{p})|b^\dagger(\vec{k})b(\vec{k})|q(\vec{p})\rangle &=(2k^+)(2\pi)^{D-1}\delta^{(D-1)}(\vec{k}-\vec{p})(2p^+)(2\pi)^{D-1}\delta^{(D-1)}(\vec{k}-\vec{p}),\\
    \langle q(\vec{k}\,'')g(\vec{l}\,')|b^\dagger(\vec{k})b(\vec{k})|q(\vec{k}\,')g(\vec{l})\rangle&=(2l^+)(2\pi)^{D-1}\delta^{(D-1)}(\vec{l}-\vec{l}\,')(2k^+)(2\pi)^{D-1}\delta^{(D-1)}(\vec{k}-\vec{k}\,'')\\
    &\times(2k'^+)(2\pi)^{D-1}\delta^{(D-1)}(\vec{k}\,'-\vec{k}).
\end{aligned}
\end{equation}
Note how the longitudinal momenta in these expressions cancel with the integral measures such that there is a factor of $1/(2p^+2l^+)$ left in the second term on the last row of Eq.~\eqref{eq: quark in quark PDF expansion}.

Expanding the renormalization constant and the wave function perturbatively in the bare (dimensionless) strong coupling $g$ as
\begin{align}
    Z_q&=1+g^2Z_{q}^{(1)}+g^4Z_{q}^{(2)}+\mathcal{O}(g^6),\\
    \psi^{q\rightarrow qg}&=g \psi^{q\rightarrow qg}_{(0)}+g^3\psi^{q\rightarrow qg}_{(1)}+\mathcal{O}(g^5),
\end{align}
the quark-in-quark PDF in Eq.~\eqref{eq: quark in quark PDF expansion} becomes up to $\mathcal{O}(g^2)$:
\begin{equation}\label{Eq: PDF expansion LO}
    f^0_{q/q}(\mu,x)=\delta(1-x)+g^2\left[ Z_{q}^{(1)}(p^+)\delta(1-x)+\frac{\bar\mu^{4-D}}{16\pi(p^+)^2x(1-x)}\int \frac{\mathrm{d}^{D-2}\mathbf{k}}{(2\pi)^{D-2}}|\psi^{q\rightarrow qg}_{(0)}(\vec{p},\vec{k},\vec{p}-\vec{k})|^2\right]+\mathcal{O}(g^4).
\end{equation}
The lowest order contribution is just a delta-function as it should, indicating that our normalization is correct.

\begin{figure}
    \centering
    \includegraphics[width=0.65\linewidth]{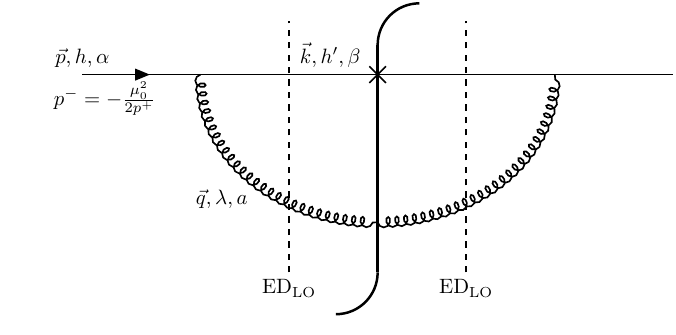}
    \caption{Leading order diagram contributing to the quark-quark splitting function, built from the square of $\psi^{q\rightarrow qg}_{(0)}$. Here, the dashed lines represent the intermediate states (and the corresponding energy denominators), the curvy line in the middle separates the wave function from its conjugate, and the cross represents that the quark is measured, i.e. it carries the momentum fraction $x$. Here $h$ and $h'$ are quark helicities, $\lambda$ is the gluon polarization, and $\alpha, \beta$ and $a$ are color indices in the fundamental and adjoint representations, respectively. The three-momenta are $\vec{p}=(p^+,\mathbf{0})$, $\vec{k}=(xp^+,\mathbf{k})$, and $\vec{q}=\vec{p}-\vec{q}=((1-x)p^+,-\mathbf{k})$. We have also indicated that the incoming quark is set off-shell with $p^-=-\mu^2_0/(2p^+)$.}
    \label{fig: LO diagram}
\end{figure}

Next, we examine the $\mathcal{O}(g^2)$ term. The square of the LCWF is depicted in Fig.~\ref{fig: LO diagram}, in which we also introduce the notation used for all diagram calculations.
To separate UV and  collinear divergences, we regulate the collinear divergences by choosing the incoming quark to have energy 
\begin{equation}
    p^-=-\frac{\mu_0^2}{2p^+},
\end{equation}
i.e. it has zero transverse momentum and a
spacelike off-shell invariant mass $m^2=-\mu_0^2$. This regulator is not only simple, but also has the agreeable physical interpretation as the potential energy obtained by the quark from being bound inside the hadron. The leading order wave function is then given by 
\begin{equation}
    g\psi^{q\rightarrow qg}_{(0)}=\frac{-gt^a_{\beta\alpha}\left[\bar u_{h'}(k)\slashed\varepsilon^*_\lambda(q)u_h(p) \right]}{\text{ED}_\text{LO}},
\end{equation}
where the energy denominator is
\begin{equation}
    \text{ED}_\text{LO}=p^--(k^-+q^-)=-\frac{1}{2p^+}\left(\frac{\mathbf{k}^2}{x(1-x)}+\mu_0^2\right).
\end{equation}
Taking the square and summing over quantum numbers (except the color of the incoming quark, $\alpha$, as it is fixed) gives
\begin{equation}\label{eq: LO WF squared}
    |g\psi^{q\rightarrow qg}_{(0)}|^2=\frac{g^2C_F(2p^+)^2x(1-x)^2\mathbf{k}^2}{\left[\mathbf{k}^2+x(1-x)\mu_0^2\right]^2}\left[ D-3+ \left(1-\frac{2}{1-x}\right)^2\right],
\end{equation}
where $C_F=(N_c^2-1)/(2N_c)$ and $N_c$ is the number of colors. The renormalization constant can be evaluated from the normalization requirement ${}_\text{int}\langle q|q\rangle_\text{int}=\langle q | q\rangle $, which at this order comprises just integrating over the $qg$ wave function, yielding
\begin{equation}
    g^2Z_{q}^{(1)}=-\frac{g^2C_F}{4\pi}\int_0^1\mathrm{d}z(1-z)\left[ D-3+ \left(1-\frac{2}{1-z}\right)^2\right]\bar\mu^{4-D}\int\frac{\mathrm{d}^{D-2}\mathbf{k}}{(2\pi)^{D-2}}\frac{\mathbf{k}^2}{\left[\mathbf{k}^2+z(1-z)\mu_0^2\right]^2}.
\end{equation}
Substituting these expressions back into the PDF in Eq.~\eqref{Eq: PDF expansion LO}, evaluating the $\mathbf{k}$-integrals, and expanding in $\varepsilon$, we find 
\begin{equation}
\begin{aligned}
    f^0_{q/q}(\mu,x) &= \delta(1-x)+\frac{\as C_F}{2\pi}\left\{ \left[\frac{1}{\varepsilon}+\ln\left(\frac{\mu^2}{x(1-x)\mu_0^2}\right) \right]\frac{1+x^2}{1-x}-2\frac{1-x+x^2}{1-x}\right\}\\
    &-\delta(1-x)\frac{\as C_F}{2\pi}\int_0^1\mathrm{d}z\left\{ \left[\frac{1}{\varepsilon}+\ln\left(\frac{\mu^2}{z(1-z)\mu_0^2}\right) \right]\frac{1+z^2}{1-z}-2\frac{1-z+z^2}{1-z}\right\}+\mathcal{O}(\as ^2),
\end{aligned}
\end{equation}
with $\as\equiv g^2/(4\pi)$.

We are now ready to renormalize the PDF to remove the UV divergence in the form of the $1/\varepsilon$ pole. Here it is essential that we put the incoming particle off shell, so that the only divergence in the bare distribution is a UV one. We do this by defining a renormalized quark PDF by subtracting the UV divergence  and the associated logarithm of the regularization scale $\mu$ as
\begin{equation}
\begin{split}
f_{q/q}(\mu_F,x) = f^0_{q/q}(\mu,x) & - \frac{\as C_F}{2\pi}\left\{ \left[\frac{1}{\varepsilon}+\ln\left(\frac{\mu^2}{\mu_F^2}\right) \right]\frac{1+x^2}{1-x} + \dots \right\}
\\
& +
\delta(1-x)\frac{\as C_F}{2\pi}\int_0^1\mathrm{d}z\left\{ \left[\frac{1}{\varepsilon}+\ln\left(\frac{\mu^2}{\mu_F^2}\right) \right]\frac{1+z^2}{1-z}+ \dots\right\},
\end{split}    
\end{equation}
where the $\dots$ denote renormalization scheme dependent finite terms. The renormalized PDF is now UV finite, but depends on a factorization scale $\mu_F$. The factorization scale should be taken as a hard scale in the scattering process, to eliminate large logarithms in the cross section.  Here we will not, however, focus on the calculation of a scattering process, but rather want to proceed towards the DGLAP evolution of the PDF. 
To obtain the DGLAP splitting kernel, we differentiate the renormalized PDF  with respect to $\ln\mu_F^2$:
\begin{equation}\label{Eq: DGLAP equation LO}
\begin{aligned}
    \frac{\mathrm{d}f_{q/q}(\mu_F,x)}{\mathrm{d}\ln\mu_F^2}&=\frac{\as C_F}{2\pi}\frac{1+x^2}{1-x}-\frac{\as C_F}{2\pi}\int_0^1\mathrm{d}z\frac{1+z^2}{1-z}\delta(1-x)+\mathcal{O}(\as ^2)\\
    &=\frac{\as }{2\pi}\int_x^1\frac{\mathrm{d}z}{z}\left[C_F\frac{1+z^2}{1-z}\right]\delta(1-\frac{x}{z})-\frac{\as }{2\pi}\int_0^1\mathrm{d}z\left[C_F\frac{1+z^2}{1-z}\right]\delta(1-x)+\mathcal{O}(\as ^2)\\
    &=\frac{\as }{2\pi}\int_x^1\frac{\mathrm{d}z}{z}\left\{C_F\left[\frac{1+z^2}{(1-z)_+}+\frac{3}{2}\delta(1-z)\right]\right\}f_{q/q}\left(\mu_F,\frac{x}{z}\right)+\mathcal{O}(\as ^2),
\end{aligned}
\end{equation}
where on the last line we noted that at zeroth order the PDF is a delta-function, with higher order corrections being $\mathcal{O}(\as ^2)$, and used the plus-prescription
\begin{equation}  \int_x^1\mathrm{d}z\frac{f(z)}{(1-z)_+}\equiv \int_x^1\frac{\mathrm{d}z}{1-z}\left[f(z)-f(1)\right]+f(1)\ln(1-z).
\end{equation}
This allows us to identify the LO quark-quark splitting function
\begin{equation}
    P^{(0)}_{qq}=C_F\left[\frac{1+z^2}{(1-z)_+}+\frac{3}{2}\delta(1-z)\right],
\end{equation}
so that Eq.~\eqref{Eq: DGLAP equation LO} takes the form of the DGLAP evolution equation \cite{Gribov:1972ri, Altarelli:1977zs, Dokshitzer:1977sg} for non-singlet distributions 
\begin{equation}\label{Eq: DGLAP equation}
   \frac{\mathrm{d}f(\mu_F,x)}{\mathrm{d}\ln\mu_F^2}=\frac{\as }{2\pi}P(\as ,x)\otimes f(\mu_F,x).
\end{equation}
Here $\otimes$ indicates the convolution integral over the longitudinal momentum fractions,
\begin{equation}
    (f\otimes g)(x)\equiv\int_x^1\frac{\mathrm{d}z}{z}f(z)g\left(\frac{x}{z}\right),
\end{equation}
and the strong coupling expansion for the splitting function is
\begin{equation}
    P(\as ,x)=P^{(0)}(x)+\frac{\as }{2\pi}P^{(1)}(x)+\mathcal{O}(\as ^2).
\end{equation}

The leading order DGLAP evolution thus arises naturally in LCPT from the UV divergence in the expectation value of the parton number operator in the hadronic state. Note that while the splitting functions are defined as the coefficients of $\ln\mu_F^2$, these logarithms are always accompanied by the $1/\varepsilon$-poles. Therefore, to simplify higher order splitting function calculations, we assume that the DGLAP evolution equation \eqref{Eq: DGLAP equation} has been established, set $\mu_0^2=\mu_F^2\equiv\mu^2$ to cancel the logarithms, and only look for the coefficient of the 1/$\varepsilon$-pole.

\section{Beyond leading order}\label{sec:nlo}
The methodology presented in the previous section is readily extended to further orders in $\as $. However, beyond leading order, antiquarks start to appear in the quark distribution due to $q\bar{q}$-mixing. To simplify the DGLAP evolution equation, we consider the non-singlet distribution
\begin{equation}
    f^{0}_{-/q}\equiv f^0_{q/q}-f^0_{\bar{q}/q}= \frac{1}{4\pi x} \int\frac{\mathrm{d}^{D-2}\mathbf{k}}{(2\pi)^{D-2}}\frac{{}_\mathrm{int}\langle q(p^+)|b^\dagger(xp^+,\mathbf{k})b(xp^+,\mathbf{k})-d^\dagger(xp^+,\mathbf{k})d(xp^+,\mathbf{k})|q(p^+)\rangle_\mathrm{int}}{{}_\mathrm{int}\langle q(p^+)|q(p^+)\rangle_\mathrm{int}},
\end{equation}
where $d^\dagger$ and $d$ are the antiquark creation and annihilation operators. Gluons treat quarks and antiquarks equally. In terms of the splitting functions this means that $P_{qq} = P_{\bar{q}\bar{q}}$, $P_{q\bar{q}} = P_{\bar{q}q}$, $P_{qg} = P_{\bar{q}g}$ and   $P_{gq} = P_{g\bar{q}}$. This leads to the evolution of  the valence quark $f_{-/h}$ decoupling from gluon distribution, with an evolution equation given by
\begin{equation}\label{Eq:singletDGLAP}
    \frac{\mathrm{d}f_{-/h}(\mu_F,x)}{\mathrm{d}\ln\mu_F^2}=\frac{\as }{2\pi}P^-(\as ,x)\otimes f_{-/h}(\mu_F,x),
\end{equation}
with the nonsinglet splitting function $P^{-}\equiv P_{qq}-P_{\bar{q}q}$ \cite{Curci:1980uw, Ellis:1996nn}.

\begin{figure}[tb!]
    \centering
    \includegraphics[width=1\linewidth]{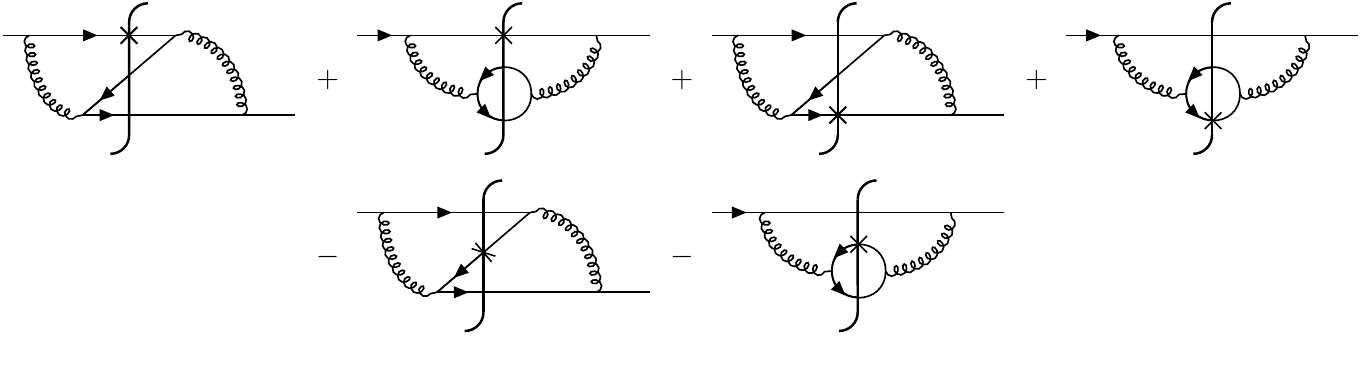}
    \caption{Diagrammatic visualization of terms arising from the expectation value of the fermion number operator in the $qq\bar{q}$ state, as given in Eq.~\eqref{eq: quark in quark PDF expansion NLO}. The different diagrams arise by contracting the tree-level LCWFs with different time and momentum orderings, with the quark carrying the momentum fraction $x$ tagged by a cross. The first and third diagrams are equal, giving a factor of two, while the fourth and sixth diagrams, where either the sea quark or sea antiquark is measured, are equal and cancel upon subtraction. Here we have not drawn diagrams with instantaneous interactions and conjugate diagrams, as well as the the multiplicative coefficients arising from summing and contracting the LCWFs.}
    \label{fig: quark measurement examples}
\end{figure}

Following a similar procedure as at leading order, we can write down further contributions to the non-singlet quark-in-quark PDF that appear at NLO  in terms of LCWFs as 
\begin{equation}\label{eq: quark in quark PDF expansion NLO}
\begin{aligned}
    &f^{0}_{-/q}(\mu,x) =Z_q(p^+)\Bigg[\delta(1-x)+\frac{1}{16\pi(p^+)^2x(1-x)}\int \frac{\mathrm{d}^{D-2}\mathbf{k}}{(2\pi)^{D-2}}\left|\psi^{q\rightarrow qg}(\vec{p},\vec{k},\vec{p}-\vec{k})\right|^2\\
    &+\frac{1}{128\pi^2 (p^+)^2x}\int \frac{\mathrm{d}^{D-2}\mathbf{k}}{(2\pi)^{D-2}}\frac{\mathrm{d}^{D-2}\mathbf{l}}{(2\pi)^{D-2}}\frac{\mathrm{d}\xi}{\xi\left(1-x-\xi\right)}\bigg\{\left|\psi^{q\rightarrow qgg}(\vec{p},\vec{k},\vec{l},\vec{p}-\vec{k}-\vec{l})\right|^2\\
    &+\left|\psi^{q\rightarrow qq\bar{q}}(\vec{p},\vec{k},\vec{l},\vec{p}-\vec{k}-\vec{l})\right|^2+\left|\psi^{q\rightarrow qq\bar{q}}(\vec{p},\vec{l},\vec{k},\vec{p}-\vec{k}-\vec{l})\right|^2-\left|\psi^{q\rightarrow qq\bar{q}}(\vec{p},\vec{l},\vec{p}-\vec{k}-\vec{l},\vec{k})\right|^2\bigg\}+\dots\Bigg],
\end{aligned}
\end{equation}
where we defined the longitudinal momentum fraction of the additional final state particle as $\xi\equiv l^+/p^+$. Here the different momentum orderings of $\psi^{qq\bar q}$ on the last line arise from evaluating the expectation value of the number operator in the $qq\bar q$-state, and correspond to measuring the first quark, the second quark, and the antiquark, respectively, where by measuring we mean specifying which parton carries the momentum $\vec{k}=(xp^+,\mathbf{k})$. Note also that the LCWFs are a sum of all distinct time-ordered diagrams, and contracting them in all possible ways gives the diagrams contributing to the splitting functions. The diagrams corresponding to different quark measurements for the $\psi^{qq\bar{q}}$ terms at $\mathcal{O}(g^4)$ are illustrated in Fig.~\ref{fig: quark measurement examples}. This figure also illustrates how, at this order, contributions that could originate from an incoming gluon state, i.e. sea quark contributions, cancel in the non-singlet distribution. Therefore the evolution kernel of the non-singlet distribution at NLO is given purely by the nonsinglet splitting function $P^-=P_{qq}-P_{\bar{q}q}$.

Expanding in powers of $g$, Eq.~\eqref{eq: quark in quark PDF expansion NLO} becomes up to $\mathcal{O}(g^4)$:
\begin{equation}
\begin{aligned} \label{eq: quark in quark PDF g expansion NLO}
    &f^0_{-/q}(\mu,x)=\delta(1-x)+g^2\left[ Z_{q}^{(1)}(p^+)\delta(1-x)+\frac{\bar\mu^{4-D}}{16\pi(p^+)^2x(1-x)}\int \frac{\mathrm{d}^{D-2}\mathbf{k}}{(2\pi)^{D-2}}\left|\psi^{q\rightarrow qg}_{(0)}(\vec{p},\vec{k},\vec{p}-\vec{k})\right|^2\right]\\
    &+g^4\Bigg[  Z_{q}^{(2)}(p^+)\delta(1-x)+ Z_{q}^{(1)}(p^+)\frac{\bar\mu^{4-D}}{16\pi(p^+)^2x(1-x)}\int \frac{\mathrm{d}^{D-2}\mathbf{k}}{(2\pi)^{D-2}} \left|\psi^{q\rightarrow qg}_{(0)}(\vec{p},\vec{k},\vec{p}-\vec{k})\right|^2\\
    &+\frac{\bar\mu^{4-D}}{16\pi(p^+)^2x(1-x)}\int \frac{\mathrm{d}^{D-2}\mathbf{k}}{(2\pi)^{D-2}}2\mathrm{Re}\left(\psi^{q\rightarrow qg}_{(1)}(\vec{p},\vec{k},\vec{p}-\vec{k})\left(\psi^{q\rightarrow qg}_{(0)}(\vec{p},\vec{k},\vec{p}-\vec{k})\right)^\dagger \right)\\
    & +\frac{\left(\bar\mu^{4-D}\right)^2}{128\pi^2 (p^+)^2x}\int \frac{\mathrm{d}^{D-2}\mathbf{k}}{(2\pi)^{D-2}}\frac{\mathrm{d}^{D-2}\mathbf{l}}{(2\pi)^{D-2}}\frac{\mathrm{d}\xi}{\xi\left(1-x-\xi\right)}\bigg\{\left|\psi^{q\rightarrow qgg}_{(0)}(\vec{p},\vec{k},\vec{l},\vec{p}-\vec{k}-\vec{l})\right|^2\\
    &+\left|\psi^{q\rightarrow qq\bar{q}}_{(0)}(\vec{p},\vec{k},\vec{l},\vec{p}-\vec{k}-\vec{l})\right|^2+\left|\psi^{q\rightarrow qq\bar{q}}_{(0)}(\vec{p},\vec{l},\vec{k},\vec{p}-\vec{k}-\vec{l})\right|^2-\left|\psi^{q\rightarrow qq\bar{q}}_{(0)}(\vec{p},\vec{l},\vec{p}-\vec{k}-\vec{l},\vec{k})\right|^2\bigg\}\Bigg]+\mathcal{O}(g^6) .
\end{aligned}
\end{equation}
Here $g$ is the bare coupling, which needs to be renormalized by absorbing into it the UV divergences emerging from the integrals in $f_{-/h}$.

At NLO accuracy we must be more careful with the renormalization procedure. As discussed in more detail in Appendix~\ref{app:renormalization}, the PDFs defined as operator expectation values are renormalized multiplicatively, and the procedure reduces to a subtraction of divergent terms in the way we presented it in the LO case only after a proper expansion in powers of $\as$. In practical terms this means that we must properly treat the subtraction of two iterations of the leading order renormalization. These contributions appear at the same order in $\as$ as the NLO contributions to the splitting function, but also include a higher power of $1/\varepsilon$ and the associated logarithm. This also introduces a scheme dependence into the definition, resulting from the possibility of performing, in addition to the $1/\varepsilon$ and logarithm, a finite renormalization of the PDF at leading order. 
Leaving a more detailed discussion of this to Appendix~\ref{app:renormalization}, we just state here that in terms of the splitting functions, the pole structure of the bare distribution $f^0_{-/q}$ 
expanded to the order in  $\as $ needed here, reads
\begin{multline}\label{eq: pdf in terms of splitting functions}
  f^0_{-/q(\mu_0)}(\mu=\mu_0,x) = \delta(1-x) + \left(\frac{\as}{2\pi}\right)
  \left[\frac{1}{\varepsilon} P^{-,(0)}(x) + C^{(0)}_{-/q}(x)\right]
+ \frac{1}{2}\left(\frac{\as}{2\pi}\right)^2 
\left[\frac{1}{\varepsilon} P^{-,(1)}(x)
+ C^{(1)}_{-/q}(x)\right]
\\  
+ \frac{1}{2}\left(\frac{\as}{2\pi}\right)^2 
\left[\frac{1}{\varepsilon}P^{-,(0)} + C^{(0)}_{-/q}\right] \otimes
\left[\frac{1}{\varepsilon}P^{-,(0)} + C^{(0)}_{-/q}\right](x)
+ \mathcal{O}(\as^3).
\end{multline}
Here, we have set all the scales to be equal $\mu_0=\mu=\mu_F$, so that the logarithms are zero. The factor of $1/2$ in front of $P^{-,(1)}$ in Eq.~\eqref{eq: pdf in terms of splitting functions}
comes from the fact that the splitting function is in fact the coefficient of $\ln\mu^2$, which at two loop order is twice the coefficient of $1/\varepsilon$. This results from the fact that a 2-loop integral is made dimensionless by a factor $\mu^{4\varepsilon}$, and $ \mu^{4\varepsilon}/\varepsilon = 1/\varepsilon + 2 \ln \mu^2 + \mathcal{O}(\varepsilon)$. Comparing the expansions \eqref{eq: quark in quark PDF g expansion NLO} and \eqref{eq: pdf in terms of splitting functions}, we see that to obtain the NLO quark-quark splitting function $P^{-,(1)}$, we need to evaluate the coefficient of the $\as ^2/((2\pi)^22\varepsilon)$-term from the integrals over the wave functions in Eq.~\eqref{eq: quark in quark PDF g expansion NLO}.  From this $1/\varepsilon$ part we have to then subtract the convolution between the leading order splitting function  $P^{-,(0)}$ and the finite part of the leading order bare distribution $C^{(0)}_{-/q}$, i.e. add a matching term
\begin{equation} \label{eq:match}
    f^\text{matching} = - \left(\frac{\as}{2\pi}\right)^2 
\frac{1}{\varepsilon}P^{-,(0)} \otimes C^{(0)}_{-/q}(x).
\end{equation}
The interpretation of this matching term is discussed in more detail in Appendix~\ref{app:renormalization}.

\begin{figure}[tb!]
    \centering
    \includegraphics[width=1\linewidth]{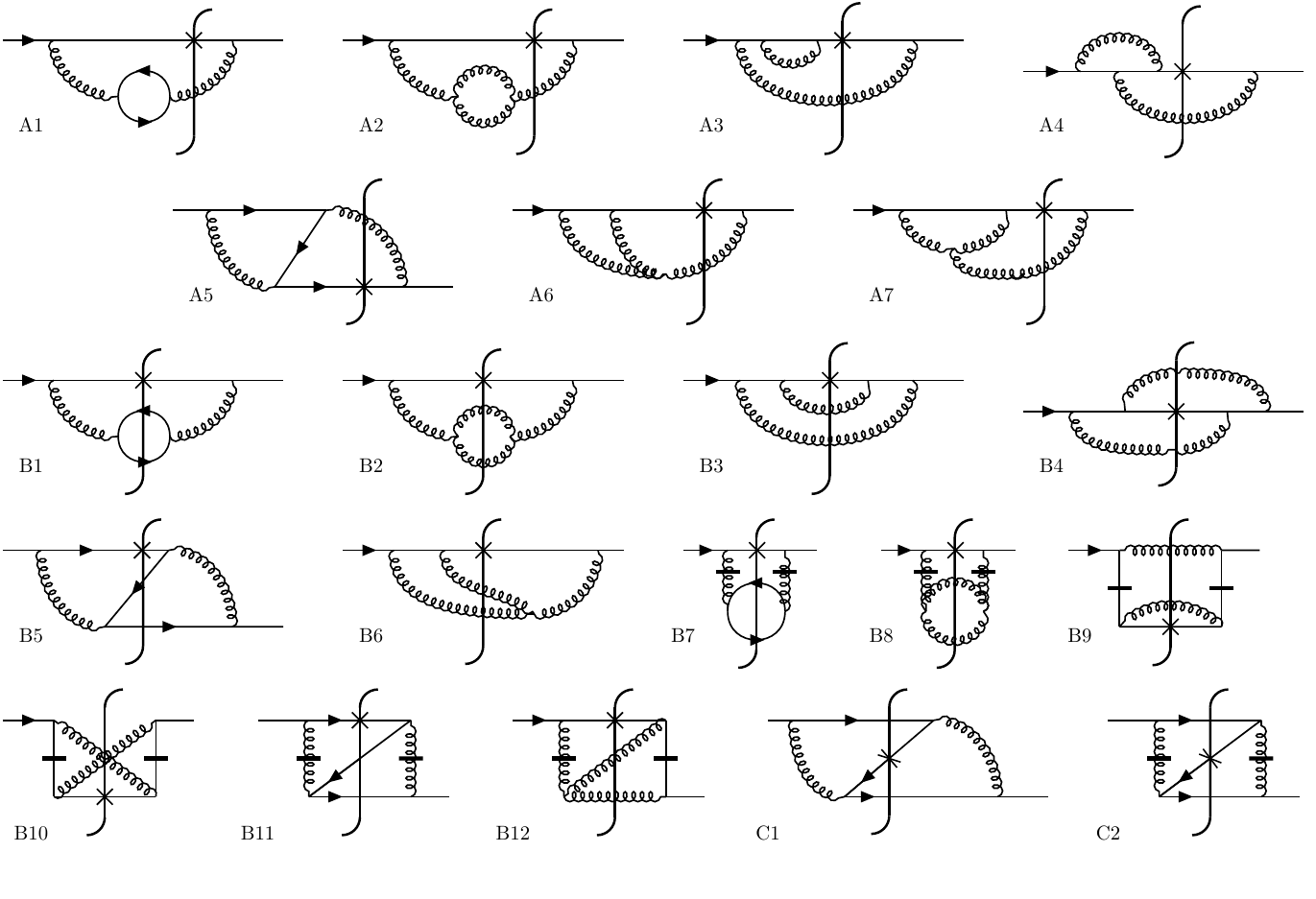}
    \caption{Diagrams contributing to the NLO non-singlet splitting function. Conjugate diagrams are not drawn explicitly, but should be included for each non-symmetric diagram. Lines with a small horizontal bar depict instantaneous interactions.}
    \label{fig: all diagrams}
\end{figure}

All the diagrams contributing to the calculation are shown in Fig.~\ref{fig: all diagrams}, where we have categorized the diagrams into
\begin{itemize}
    \item A: one-loop corrections to the $qg$ wave function times the LO wave function,
    \item B: squares of the two-particle real emission LCWFs, where a quark is measured,
    \item C: squares of the quark-antiquark emission LCWFs, where the antiquark is measured.
\end{itemize}
The sum of diagrams of type A and B give $P_{qq}^{V,(1)}$, while type C gives $P_{q\bar q}^{V,(1)}$.

It should be noted that there are many more diagrams with instantaneous interactions, and we have only drawn the ones that do not vanish trivially from the following observation: As instantaneous vertices carry no dependence of the transverse loop momenta in the numerator, diagrams with one instantaneous vertex and one ordinary one are linear in the transverse momenta, vanishing upon integration in dimensional regularization. Thus, in our case, only the diagrams with either zero or two instantaneous vertices contribute. Diagrams with two instantaneous vertices can clearly be present only for the real emissions of two particles.

In addition to these diagrams, and the subtraction term in Eq.~\eqref{eq:match}, which is related to the collinear one-loop subdivergence of the ladder diagram B3, the PDF gets contributions from the quark wave function renormalization constants and the UV renormalization of the coupling. These contributions are explained in detail below.

\section{NLO splitting function calculation}\label{sec:details}
In this section we evaluate the full NLO splitting function $P^-$ using LCPT. In the following diagram calculations we use the same notation and definitions introduced in the LO case in section \ref{section: LO derivation}. This includes regulating divergences in conventional dimensional regularization, and separating collinear and UV divergences by setting the initial quark energy to $p^-=-\mu^2/(2p^+)$. In addition, in the longitudinal integrals we encounter soft divergences, which we regulate with a cut-off $\alpha$.

The calculation is organized as follows: First, we present a detailed example calculation for a single diagram, which showcases how the NLO splitting function arises from LCWFs, and how two-loop diagrams are evaluated in LCPT. We present an automatable algorithm for the diagram calculations, with which we evaluate the other diagrams, leaving details to the appendices. We then discuss the remaining contributions from the subtraction term of the collinear subdivergence, UV renormalization and the quark wave function renormalization constants.

\subsection{\label{Section: Detailed example}Detailed example of a two-loop diagram calculation}

\begin{figure}
    \centering
    \includegraphics[width=0.75\linewidth]{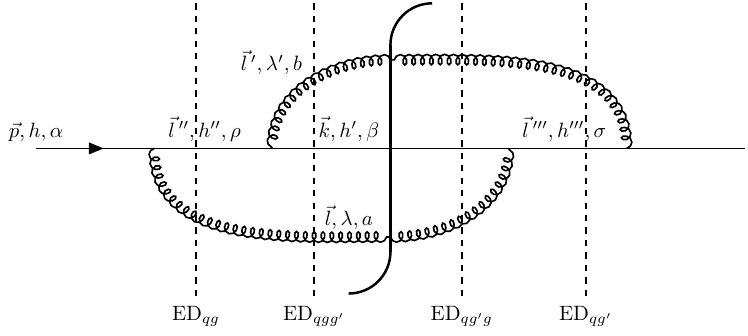}
    \caption{Diagram B4, with kinematics and energy denominators. The left hand side of the cut depicts $\psi_{(0)}^{q\rightarrow qgg'}$, while the right hand side is $\left(\psi_{(0)}^{q\rightarrow qg'g}\right)^\dagger$. The gluon carrying momentum $\vec{l}\,'=\vec{p}-\vec{k}-\vec{l}$ is tagged with a prime in the notation for the wave functions and energy denominators.}
    \label{fig: diagram 1a}
\end{figure}

As a demonstrative example, we evaluate in detail Diagram B4, as it includes the general features of the two-loop diagrams, including the most complicated integral structure, but is otherwise simple as it has no subdivergences. The diagram and its kinematics are shown in Fig.~\ref{fig: diagram 1a}.
To distinguish the two different orderings of the $qgg$ wave functions, we tag one gluon with a prime. Using the LCPT rules, the wave functions read 
\begin{align}
    \psi_{(0)}^{q\rightarrow qgg'}&=\int \measure{l''}(2\pi)^{D-1}\delta^{(D-1)}(\vec{p}-\vec{l}-\vec{l}\,'')\frac{(-gt^b_{\beta\rho})\Big[\Bar{u}_{h'}(k)\slashed{\varepsilon}^*_{\lambda'}(l')u_{h''}(l'')\Big](-gt^a_{\rho\alpha})\Big[\Bar{u}_{h''}(l'')\slashed{\varepsilon}^*_{\lambda}(l)u_h(p)\Big]}{\text{ED}_{qg}\text{ED}_{qgg'}},\label{Eq: WF qgg 1}\\
    %%%
    \psi_{(0)}^{q\rightarrow qg'g}&=\int \measure{l'''}(2\pi)^{D-1}\delta^{(D-1)}(\vec{l}\,'''-\vec{k}-\vec{l}\,)\frac{(-gt^a_{\beta\sigma})\Big[\Bar{u}_{h'}(k)\slashed{\varepsilon}^*_{\lambda}(l)u_{h'''}(l''')\Big](-gt^b_{\sigma\alpha})\Big[\Bar{u}_{h'''}(l''')\slashed{\varepsilon}^*_{\lambda'}(l')u_h(p)\Big]}{\text{ED}_{qg'}\text{ED}_{qg'g}},\label{Eq: WF qgg 2}
\end{align}
where the energy denominators are:
\begin{align}
    \text{ED}_{qg}&=\frac{-1}{2p^+\xi(1-\xi)}\left(\mathbf{l}^2+\xi(1-\xi)\mu^2\right),\\
    \text{ED}_{qgg'}&=\text{ED}_{qg}-\frac{(1-\xi)}{2p^+x(1-x-\xi)}\Big(\mathbf{k}+\frac{x}{1-\xi}\mathbf{l}\Big)^2,\\
    \text{ED}_{qg'}&=\frac{-1}{2p^+(x+\xi)(1-x-\xi)}\left((\mathbf{k}+\mathbf{l})^2+(x+\xi)(1-x-\xi)\mu^2\right),\\
    \text{ED}_{qg'g}&=\text{ED}_{qg'}-\frac{\xi}{2p^+x(x+\xi)}\Big(\mathbf{k}-\frac{x}{\xi}\mathbf{l}\Big)^2.
\end{align}

According to Eq.~\eqref{eq: quark in quark PDF g expansion NLO}, the contribution from Diagram B4 is 
\begin{equation}
    f^-_{\text{B4}}= 2\frac{\bar\mu^{4-D}}{128\pi^2x(p^+)^2}\int\frac{\mathrm{d}^{D-2}\textbf{k}}{(2\pi)^{D-2}}\frac{\mathrm{d}^{D-2}\textbf{l}}{(2\pi)^{D-2}}\frac{\mathrm{d}\xi}{\xi\left(1-x-\xi\right)}\psi_{(0)}^{q\rightarrow qgg'}\left(\psi_{(0)}^{q\rightarrow qg'g}\right)^\dagger,
\end{equation}
where the factor of two comes from the identical contribution of the hermitian conjugate of this diagram. Substituting in the wave functions \eqref{Eq: WF qgg 1} and \eqref{Eq: WF qgg 2}, and evaluating the color algebra and the integrals over delta functions, we get 
\begin{equation}\label{Eq: Ia start}
\begin{aligned}
    f^-_{\text{B4}}&=\as ^2C_F\left(C_F-\frac{C_A}{2}\right)\int_{\alpha}^{1-x-\alpha}\mathrm{d}\xi\frac{(1-x-\xi)^2(x+\xi)}{(1-\xi)}\bar\mu^{4-D}\int \frac{\mathrm{d}^{D-2}\mathbf{l}}{(2\pi)^{D-2}}\frac{1}{\mathbf{l}^2+\xi(1-\xi)\mu^2}\\
    &\times \bar\mu^{4-D}\int\frac{\mathrm{d}^{D-2}\mathbf{k}}{(2\pi)^{D-2}}\frac{[\bar u_{h}(p)\slashed \varepsilon_{\lambda'}(l') u_{h'''} (l''')][\bar u_{h'''}(l''')\slashed\varepsilon_{\lambda}(l)u_{h'}(l)][\bar u_{h'}(k)\slashed\varepsilon^*_{\lambda'}(l')u_{h''}(l'')][\bar u_{h''}(l'')\slashed\varepsilon^*_{\lambda}(l)u_{h}(p)]}{(\mathbf{k}+\mathbf{l})^2\left[(\mathbf{k}+\mathbf{l})^2+\frac{\xi(1-x-\xi)}{x}(\mathbf{k}-\frac{x}{\xi}\mathbf{l})^2\right]\left[(\mathbf{k}+\frac{x}{1-\xi}\mathbf{l})^2+\frac{x(1-x-\xi)}{\xi(1-\xi)^2}\mathbf{l}^2\right]},
\end{aligned}
\end{equation}
where $C_A=N_c$, and we set $\mu^2=0$ in the first integral since it is finite in the collinear limit.

The two-loop integral in Eq.~\eqref{Eq: Ia start} is of the most difficult kind encountered in the computation of the NLO splitting functions, and the systematic evaluation of such integrals is one of the main results of this paper. Let us therefore review the evaluation of the two-loop integrals by considering a general case:
\begin{equation}\label{Eq: two-loop integral example}
    \mathcal{I}=\left(\bar\mu^{4-D}\right)^2\int \frac{\mathrm{d}^{D-2}\mathbf{l}}{(2\pi)^{D-2}}\frac{1}{\mathbf{l}^2+\Delta_0\mu^2} \int\frac{\mathrm{d}^{D-2}\mathbf{k}}{(2\pi)^{D-2}}\frac{\mathcal{N}}{\mathbf{k}^2\left[\mathbf{k}^2+\Delta_1(\mathbf{k}+\Delta_2\mathbf{l})^2\right]\left[(\mathbf{k}+\Delta_3\mathbf{l})^2+\Delta_4\mathbf{l}^2\right]}.
\end{equation}
Here $\mathcal{N}$ stands for the numerator structure, and $\Delta_i$ are functions of the longitudinal momentum fractions only, i.e. they are independent of the loop momenta.
The evaluation of these integrals requires automatization with computer tools, and Eq.~\eqref{Eq: two-loop integral example} is the starting point of our semi-automatic framework. The steps of the evaluation process are as follows:
\begin{enumerate}
    \item \textbf{Evaluation of spinor algebra:} The numerator $\mathcal{N}$ consists of Dirac spinors and polarization vectors, such as in Eq.~\eqref{Eq: Ia start}. These are evaluated using the methodology introduced in Ref.~\cite{Hanninen:2017ddy}, where the spinor structure is decomposed into its symmetric ($\delta^{ij}$) and anti-symmetric ($[\gamma^i,\gamma^j]$) parts, helicities are summed over using the spinor sum relations, and remaining gamma-matrices are contracted. For the general two-loop case, the numerator is too cumbersome to be evaluated by hand, but the benefit of the helicity basis method is that it can be fully automated with e.g. FORM. In general, the result of the numerator contractions is
    \begin{equation}
    \mathcal{I}=\left(\bar\mu^{4-D}\right)^2\int \frac{\mathrm{d}^{D-2}\mathbf{l}}{(2\pi)^{D-2}}\frac{1}{\mathbf{l}^2+\Delta_0\mu^2} \int\frac{\mathrm{d}^{D-2}\mathbf{k}}{(2\pi)^{D-2}}\frac{N_1\mathbf{k}^2\mathbf{l}^2+N_2\mathbf{k}^2(\mathbf{k}\cdot\mathbf{l})+N_3(\mathbf{k}\cdot\mathbf{l})\mathbf{l}^2+N_4(\mathbf{k}\cdot\mathbf{l})^2}{\mathbf{k}^2\left[\mathbf{k}^2+\Delta_1(\mathbf{k}+\Delta_2\mathbf{l})^2\right]\left[(\mathbf{k}+\Delta_3\mathbf{l})^2+\Delta_4\mathbf{l}^2\right]},
    \end{equation}
    where the coefficients $N_i$ are complicated functions of $x$ and $\xi$. The spinor algebra results also in terms proportional to the antisymmetric term $[\gamma^i,\gamma^j]$, but these terms must vanish after the $\mathbf{l}$-integral, so we have ignored them.
    
    \item \textbf{Tensor reduction:} The $\mathbf{k}$-integral can be performed by reducing the tensor structure to a combination of scalar integrals using the Passarino-Veltman method applied to LCPT. The tensor reduction and evaluation of scalar master integrals is presented in detail in Appendix \ref{Appendix: integrals}. In terms of the notation introduced in the Appendix, in which $\mathcal{B}$ and $\mathcal{C}$ are one-loop two-point and three-point integrals respectively, with a number of indices corresponding to the tensor rank, we have
    \begin{equation}
    \begin{aligned}
        \mathcal{I}&=\bar\mu^{4-D}\int \frac{\mathrm{d}^{D-2}\mathbf{l}}{(2\pi)^{D-2}}\frac{1}{\mathbf{l}^2+\Delta_0\mu^2} \left[N_1\mathbf{l}^2\mathcal{B}_0+N_2l^i\mathcal{B}^i+N_3\mathbf{l}^2l^i\mathcal{C}^i+N_4l^il^j\mathcal{C}^{ij}\right]\\
        &=\bar\mu^{4-D}\int \frac{\mathrm{d}^{D-2}\mathbf{l}}{(2\pi)^{D-2}}\frac{1}{\mathbf{l}^2+\Delta_0\mu^2} \left[N_1\mathbf{l}^2\mathcal{B}_0+N_2\mathbf{l}^2 \mathcal{B}_1+N_3\mathbf{l}^4\mathcal{C}_1+N_4(\mathbf{l}^4\mathcal{C}_{21}+\mathbf{l}^2\mathcal{C}_{22})\right],
    \end{aligned}
    \end{equation}
    where expressions for the scalar integrals $\mathcal{B}_1$, $\mathcal{C}_1$, $\mathcal{C}_{21}$, and $\mathcal{C}_{22}$ in terms of master integrals are derived in detail in the Appendix.

    \item \textbf{Evaluation of transverse scalar integrals:} The denominator structure in $\mathcal{I}$ differs from standard covariant theory loop integrals due to the momenta being weighted by factors of $\Delta_i$. However, with some variable changes, the remaining scalar integrals over can be brought to a form resembling covariant theory integrals that are evaluated with standard methods, see Appendix \ref{Appendix: integrals}. After performing the first loop integral, the second integral will always be of the simple form
    \begin{equation}
        \bar\mu^{4-D}\int \frac{\mathrm{d}^{D-2}\mathbf{l}}{(2\pi)^{D-2}}\frac{(\mathbf{l}^2)^{a}}{\left[\mathbf{l}^2+\Delta_0\mu^2\right]^b},
    \end{equation}
    where $a$ and $b$ are real numbers that can depend on $\varepsilon$. The result of this integral is presented in Eq.~\eqref{Eq: master integral}.

    \item \textbf{Expansion in $\varepsilon$ and evaluation of the longitudinal integral:} After the transverse integrals have been evaluated, we expand in $\varepsilon$ and perform the final elementary integration over the longitudinal momentum fraction $\xi$. These are straightforward to do with e.g. Mathematica.
\end{enumerate}

These steps are used to evaluate all two-loop integrals in this paper. Continuing with our example diagram, we start from Eq.~\eqref{Eq: Ia start} and apply the above algorithmic steps to obtain
\begin{equation}\label{Eq: diagram B4 result}
    f^-_{\text{B4}}\Big|_{1/\varepsilon}=\frac{1}{2\varepsilon}\frac{\as ^2}{(2\pi)^2}C_F\left(C_F-\frac{C_A}{2}\right)\bigg\{\frac{1+x^2}{1-x}\Big[-\ln^2x+4\ln\frac{1-x}{\alpha}\Big]+2(1+x)\ln x\bigg\}.
\end{equation}
Here, $f^-_{\text{B4}}\Big|_{1/\varepsilon}$ means that we have kept only the $1/\varepsilon$-pole, which gives the contribution to the splitting function. Note that the result of this diagram is in full agreement with the covariant theory result \cite{Curci:1980uw} for this particular topology, as the only other diagram contributing to this topology, diagram B10, vanishes (see Appendix \ref{appendix: diagram results}). In general, we find that, although LCPT has more diagrams to evaluate, the sum of all diagrams in a given topology is in one-to-one correspondence with the result of the same topology evaluated in covariant perturbation theory in Ref.~ \cite{Curci:1980uw}.

Each diagram in Fig. \ref{fig: all diagrams} is evaluated using the above methodology. Since the example calculation covers the general features of all diagrams, we do not present here explicit calculations for each diagram, instead the results for each diagram are given in Appendix \ref{appendix: diagram results}.

Although in this paper we do not present the finite terms as they do not contribute to the splitting functions, they are nevertheless obtained by the above procedure. The automated framework is thus readily applied to two-loop calculations where also the finite parts are needed.

\subsection{One-loop subdivergences and UV renormalization}
The $1/\varepsilon$-poles of $f^-$ obtained in the diagram calculations include subdivergences from the one-loop subdiagrams, which do not contribute to the splitting functions and should be removed. First, there are UV divergences, which are removed by absorbing them to the renormalization of the coupling constant. Therefore, we introduce the dimensionless renormalized coupling $g_R$ in Eq.~\eqref{eq: quark in quark PDF g expansion NLO} by making the replacement
\begin{equation}
    g=g_R-\frac{g_R^3}{16\pi^2}\left(\frac{11}{6}C_A-\frac{2T_FN_f}{3}\right)\frac{1}{\varepsilon}+\mathcal{O}(g_R^4),
\end{equation}
with $T_F=1/2$ and $N_f$ being the number of quark flavors. At $\mathcal{O}(g_R^4)$ this introduces the counterterm
\begin{equation}\label{Eq: counterterm}
\begin{aligned}
    f^{\text{UV}}&=-2\as ^2\left(\frac{11}{6}C_A-\frac{2T_FN_f}{3}\right)\frac{1}{\varepsilon}\frac{1}{16\pi(p^+)^2x(1-x)}\bar{\mu}^{4-D}\int \frac{\mathrm{d}^{D-2}\mathbf{k}}{(2\pi)^{D-2}}|\psi^{q\rightarrow qg}_{(0)}(\vec{p},\vec{k})|^2\\
    &=\frac{1}{2\varepsilon}\frac{\as ^2}{(2\pi)^2}C_F\left(\frac{11}{3}C_A-\frac{4T_FN_f}{3}\right)\left[\frac{1+x^2}{1-x}\ln (x(1-x))+2\frac{1-x+x^2}{1-x}\right] +\mathcal{O}\left(\frac{1}{\varepsilon^2}\right),
\end{aligned}
\end{equation}
where from here on $\as$ corresponds to the renormalized coupling.

Although it is clear that the UV poles must be absorbed into the running of the coupling, it is not immediately obvious how the correct beta-function emerges from the LCWFs, which include final state propagator corrections but not initial state propagator corrections. A detailed derivation of the one-loop QCD beta function in LCPT was performed in Ref.~\cite{Lappi:2016oup}, where a renormalized quark-gluon state
\begin{equation}
    \sqrt{Z_q(p^+)}\WFqgg{}(\vec{p},\vec{k})|q(\vec{k})g(\vec{q})\rangle=\frac{\sqrt{Z_q(p^+)}}{\sqrt{Z_q(k^+)Z_g(q^+)}}\WFqgg{}(\vec{p},\vec{k})|q(\vec{k})g(\vec{q})\rangle_R
\end{equation}
was introduced to make contact with covariant perturbation theory. Here, the denominator cancels half of the outgoing propagator correction diagrams included in $\WFqgg{}$ and the numerator introduces half of the incoming state corrections absent in $\WFqgg{}$. A similar cancellation happens in the splitting function calculation automatically: Half of the UV divergences emerging in the propagator correction diagrams A1, A2, and A3 are cancelled by UV divergences from the corresponding real emission diagrams B1, B2, B3, B7, B8, and B9 arising from the phase space integration over one of the outgoing particles. Furthermore, the one-loop renormalization constant term in \eqref{eq: quark in quark PDF g expansion NLO},
\begin{equation}\label{eq: one-loop renormalization constant contribution}
\begin{aligned}
    f^{Z_q^{(1)}}&=g^4_R Z_{q}^{(1)}(p^+) \frac{\bar\mu^{4-D}}{16\pi(p^+)^2x(1-x)}\int \frac{\mathrm{d}^{D-2}\mathbf{k}}{(2\pi)^{D-2}}\left|\psi^{q\rightarrow qg}_{(0)}(\vec{p},\vec{k})\right|^2\\
    &=- \frac{1}{2\varepsilon}\frac{\as ^2}{(2\pi)^2}C_F\left(3+4\ln\alpha\right)\left[\frac{1+x^2}{1-x}\ln (x(1-x))+2\frac{1-x+x^2}{1-x}\right]+\mathcal{O}\left(\frac{1}{\varepsilon^2}\right),
\end{aligned}
\end{equation}
introduces the UV poles corresponding to the incoming quark propagator correction. Together, these combine to give the one-loop running of the coupling, and are cancelled by the counter-term of Eq.~\eqref{Eq: counterterm}.

The ladder diagram B3 has a  subdivergence resulting from the inner one-loop contribution, leading to a double pole $1/\varepsilon^2$ for the diagram overall.  This corresponds to  the contribution of two steps of leading order evolution in Eq. \eqref{eq: pdf in terms of splitting functions} and is easily removed. There is also a related $\as ^2/\varepsilon$-term related to the convolution of the LO splitting function $P^{(0)}_{qq}$ and the finite parts of the $\mathcal{O}(\as)$ terms of $f^-$ in Eq. \eqref{eq: quark in quark PDF g expansion NLO}, coming from the integral over the squared LO wave function. As discussed above, this is removed  by adding 
the corresponding counterterm~\eqref{eq:match} 
\begin{equation}\label{eq: matching counterterm}
\begin{aligned}
    f^\text{matching}&=\frac{1}{\varepsilon}\frac{\as ^2}{(2\pi)^2}P^{(0)}_{q/q}\otimes \left[\frac{1+x^2}{1-x}\ln (x(1-x))+2\frac{1-x+x^2}{1-x}\right]\\
    &=\frac{1}{2\varepsilon}\frac{\as ^2}{(2\pi)^2}C_F^2\bigg\{\frac{1+x^2}{1-x}\left[3-\frac{2\pi^2}{3}-\ln x+11\ln(1-x)-2\ln^2x+6\ln^2(1-x)-4\ln\alpha-2\ln^2\alpha\right]\\
     &+(1-x)\left[-3-3\ln x\right]+(1+x)\left[\ln x+2\ln x\ln(1-x)+\ln^2 x+2\mathrm{Li}_2(1-x)\right]\bigg\},
\end{aligned}
\end{equation}
where $\mathrm{Li}_2(x)$ is the standard dilogarithm function. The same counterterm emerges in the original Curci-Furmanski-Petronzio method from the expansion of the collinear projection operators.

Note that both $f^{Z_q^{(1)}}$ and $f^\text{matching}$ would also include a contribution from the finite part of $Z_{q}^{(1)}$ convoluted with the LO splitting function. However, these contributions are equal but opposite sign, and cancel each other. This cancellation is required, since the finite parts of the renormalization constant depend on the arbitrary choice of the collinear regulator.

We are now ready to write down the result for the splitting function for $x\neq1$. Combining the results from diagram calculation from Appendix \ref{appendix: diagram results} with the renormalization terms \eqref{Eq: counterterm} and \eqref{eq: one-loop renormalization constant contribution}, matching them to the right hand side of Eq.~\eqref{eq: pdf in terms of splitting functions}, and adding the matching correction term \eqref{eq: matching counterterm}, we get the results for the valence splitting functions for $x\neq 1$ to be
\begin{equation}\label{Eq: x neq 1 splitting qq}
\begin{aligned}
    P_{qq}^{V,(1)}(x\neq 1)&=C_F^2\left\{-\left[2\ln x\ln(1-x)+\frac{3}{2}\ln x\right]\frac{1+x^2}{1-x}-\left(\frac{3}{2}+\frac{7}{2}x\right)\ln x-\frac{1}{2}(1+x)\ln^2x-5(1-x)\right\}\\
    &+C_FC_A\left\{\left[\frac{1}{2}\ln^2x+\frac{11}{6}\ln x+\frac{67}{18}-\frac{\pi^2}{6}\right]\frac{1+x^2}{1-x}+(1+x)\ln x+\frac{20}{3}(1-x)\right\}\\
    &+C_FN_fT_F\left\{-\left[\frac{2}{3}\ln x+\frac{10}{9}\right]\frac{1+x^2}{1-x}-\frac{4}{3}(1-x)\right\},
\end{aligned}
\end{equation}
and
\begin{equation}
\begin{aligned}
    P_{q\bar q}^{V,(1)}(x\neq 1)=C_F\left(C_F-\frac{C_A}{2}\right)\left\{\left[-4\mathrm{Li}_2(-x)+\ln^2x-4\ln x\ln(1+x)-\frac{\pi^2}{3}\right]\frac{1+x^2}{1+x}+2(1+x)\ln x+4(1-x)\right\}.
\end{aligned}
\end{equation}
As should, the soft regulator $\alpha$ cancels upon summing the diagrams.

\subsection{Endpoint contribution from quark wave function renormalization constant}
The final contribution to the splitting function comes from the two-loop quark wave function renormalization constant $Z_{q}^{(2)}$, which contributes only at the endpoint $x=1$. The renormalization constant is evaluated by summing over all possible intermediate states and squaring the corresponding LCWFs. Thus, for each diagram contributing to the non-singlet PDF in Fig.~\ref{fig: all diagrams}, there is a corresponding contribution to $(Z^{(2)}_{q})^{-1}$, however this time integrating also over the quark momentum fraction.

Note how it is specifically the non-singlet distribution -- where the contribution of the sea quarks cancel -- whose diagrams correspond to those of the renormalization constant. 
Trying to reconstruct the quark renormalization constant contribution from integrating the full the quark–in–quark distribution would double-count contributions, because then the measured quark can also originate from the gluon splitting into a quark-antiquark pair. The gluon to quark-antiquark splitting corresponding contribution to $Z_{q}^{(2)}$ is already included in diagram A1 contributing to the nonsinglet distribution and should not be included a second time. Therefore, the contribution to the endpoint from $Z_{q}^{(2)}$ can be written as
\begin{equation}\label{Eq: endpoint}
    \delta(1-x)Z_{q}^{(2)}(p^+)\Big|_{1/\varepsilon}=-\frac{1}{2\varepsilon}\frac{\as ^2}{(2\pi)^2}\delta(1-x)\int_0^{1-\alpha} \ud zP^{-,(1)}(z).
\end{equation}
This is equivalent to solving the endpoint contribution from the condition 
\begin{equation}
    \int_0^1 \ud z P^{-,(1)}(z)=0,
\end{equation}
i.e. the condition for conservation of fermion number, which is how the endpoint contribution was originally computed in Ref.~\cite{Curci:1980uw}.

Evaluating the endpoint contribution from Eq.~\eqref{Eq: endpoint} amounts to adding
\begin{equation}
    \delta(1-x)\left\{C_F^2\left[\frac{3}{8}-\frac{\pi^2}{2}+6\zeta(3)\right]+C_FC_A\left[\frac{17}{24}+\frac{11\pi^2}{18}-3\zeta(3)\right]-C_FN_fT_F\left[\frac{1}{6}+\frac{2\pi^2}{9}\right]\right\}
\end{equation}
and changing
\begin{equation}
    \frac{1}{1-x}\rightarrow \frac{1}{(1-x)_+}
\end{equation}
in Eq.~\eqref{Eq: x neq 1 splitting qq}.

\subsection{Final result}
Combining the results for $x\neq 1$ and $x=1$ from above, we get the final result for the two-loop non-singlet splitting function to be
\begin{equation}
    P^{-,(1)}(x)=P_{qq}^{V,(1)}(x)-P_{q\bar q}^{V,(1)}(x),
\end{equation}
where
\begin{equation}
\begin{aligned}
    P_{qq}^{V,(1)}(x)&=C_F^2\bigg\{\left[\frac{3}{8}-\frac{\pi^2}{2}+6\zeta(3)\right]
\delta(1-x)-\left[2\ln x\ln(1-x)+\frac{3}{2}\ln x\right]\frac{1+x^2}{(1-x)_+}\\&\qquad\quad-\left(\frac{3}{2}+\frac{7}{2}x\right)\ln x-\frac{1}{2}(1+x)\ln^2x-5(1-x)\bigg\}\\
    &+C_FC_A\bigg\{\left[\frac{17}{24}+\frac{11\pi^2}{18}-3\zeta(3)\right]\delta(1-x)+\left[\frac{1}{2}\ln^2x+\frac{11}{6}\ln x+\frac{67}{18}-\frac{\pi^2}{6}\right]\frac{1+x^2}{(1-x)_+}\\
    &\qquad\qquad+(1+x)\ln x+\frac{20}{3}(1-x)\bigg\}\\
    &+C_FN_fT_F\left\{-\left[\frac{1}{6}+\frac{2\pi^2}{9}\right]\delta(1-x)-\left[\frac{2}{3}\ln x+\frac{10}{9}\right]\frac{1+x^2}{(1-x)_+}-\frac{4}{3}(1-x)\right\},
\end{aligned}
\end{equation}
and
\begin{equation}
\begin{aligned}
    P_{q\bar q}^{V,(1)}(x)=C_F\left(C_F-\frac{C_A}{2}\right)\left\{\left[-4\mathrm{Li}_2(-x)+\ln^2x-4\ln x\ln(1+x)-\frac{\pi^2}{3}\right]\frac{1+x^2}{1+x}+2(1+x)\ln x+4(1-x)\right\}.
\end{aligned}
\end{equation}
Our result agrees perfectly with the original result of Ref.~\cite{Curci:1980uw}, demonstrating the validity of our methodology, and the agreement of LCPT and covariant theory for higher order perturbative predictions.

\section{Conclusions} \label{sec:conc}

Let us now briefly summarize the calculation that we have presented in this paper, and outline some future directions. First and foremost, we demonstrated by an explicit calculation that there are no inherent problems in extending LCPT calculations in gauge theories to higher order. The perturbative rules defining the LCWFs at higher loop orders are explicit, and lead to integrals that are calculable. Quantization in light cone time, and the axial light cone gauge condition, lead to a loss of manifest Lorentz-invariance in the intermediate expressions. To what extent this is a problem depends on the regularization method and the physical process one studies: for scattering at high energy the manifest boost invariance in the description of QCD bound states can be a more important property to preserve. In any case, in perturbative calculations rotational invariance can be systematically restored order by order in perturbation theory \cite{Burkardt:1991tj,Beuf:2021srj,Beuf:2022ndu}. At a more practical calculational level, this leads one to have to deal with two physically different kinds of integrals: transverse and longitudinal ($k^+$) momentum. The longitudinal momentum is strictly positive, and thus no ambiguities from $1/k^+$ poles that are present in covariant perturbation theory in light cone gauge arise. The corresponding divergences at $k^+\to 0$ can be regularized, e.g. with a cutoff as here, and eventually cancel from physical observables, or are absorbed into the appropriate renormalization group equations (w.g. the B-JIMWLK equation). Transverse  momentum integrals appear in a form that is slightly different than in $4-2\varepsilon$ dimensional covariant theory, because the energy denominators contain additional longitudinal momentum-fraction dependent coefficients. However, they can be computed in dimensional regularization with a relatively strightforward generalization of standard methods. 

We have tried to present the calculation of the LCWF's in a way that makes it clear that the process in this paper is inherently automatizable, which is an essential feature of modern precision calculations. Once one defines the incoming and outgoing (dressed and bare) states in the wavefunction, mathematical expressions corresponding to all the contributing diagrams can be generated by iterating over a finite number of vertices present in the theory. Numerators can be simplified with standard spinor algebra tools, and tensor integrals reduced to the evaluation of a standardizable set of scalar master integrals.
We believe that at least up to the evaluation of the transverse momentum integrals, all two-loop processes are inherently doable in LCPT. What remains to do then are integrals over the longitudinal momenta, of integrands containing rational functions and logarithms. In the case of CGC calculations one also needs to perform  Fourier transforms of the LCWFs into transverse coordinate space. Also here there is no fundamental reason why something would be undoable at two loops, although it remains to be seen in practice at what point one has to resort to numerical evaluations. In our view the next step in the program of going to two loops in LCPT is to start automatizing these steps, starting from the generation of the diagrams, the spinor algebra and the evaluation of the transverse momentum integrals. 

In QCD problems it is often advantageous to consider spacelike off-shell states. 
One might wonder whether an off-shell external parton leads to spurious gauge-dependent effects~\cite{vanHameren:2012if,vanHameren:2013csa,Bailey:2022wqy} in gauge theory. We believe that when we work in axial gauge this should not be a problem; in fact this is one of the main advantages of axial gauge. Off-shell external partons  lead to additional mass terms in the energy denominators, but significantly simplify the physical interpretation in terms of collinear divergences. Since in the DIS process an off-shell incoming photon is required by the physical process, one already has significant experience from higher order CGC  calculations in dealing with it.

As a side product of our calculation, we have provided a concise explanation of how to define and renormalize parton distribution functions in the LCPT language. In our opinion this provides a very physical and transparent way to understand QCD factorization. In covariant theory, one usually introduces the concept of a ``renormalized field'', related to the bare field by a multiplicative factor. In the context of QCD partonic dynamics this is a somewhat confusing concept, since the actual physical single particle states of the interacting theory are states with multiple  bare  particles, not multiplicatively renormalized single (bare) particle states. In LCPT, the language is different. Only the actual parameters in the Hamiltonian, the coupling constant and quark masses, are  \emph{re}normalized. There is no field renormalization, but in stead one works with normalized bare and dressed \emph{states} that are perturbatively related to each other. It is our hope that, while this is not the main focus of this paper, our presentation can also provide an alternative, perhaps more understandable, point of view to the nature of partonic distributions in QCD.

\section*{Acknowledgments}
This work has been supported by the Research Council of Finland, as a part of the Centre of Excellence in Quark Matter (TL, project 364191), the Centre of Excellence in Neutron Star Physics (RP, MS 374062 and 374063) and projects 347499,
353772, 354533, and 354572 (MS,RP). We are grateful for comments on the manuscript from Xuan-Bo Tong.

\appendix

\section{Renormalization of parton distribution functions}\label{app:renormalization}

We do not wish to present a full and detailed  discussion of QCD factorization and DGLAP evolution further here, since it evokes confusion and controversy (see e.g.~\cite{Candido:2020yat,Collins:2021vke,Candido:2023ujx}) even  after (or precisely because of) having been addressed at so much  length  in the literature. Rather, we will just state here the results that are needed for our calculation, so the discussion in this Appendix will be minimalistic. Our aim here is to justify how we arrive at \eq\eqref{eq: pdf in terms of splitting functions}. Recall that
this is the equation that we use to relate the l.h.s. that we calculate in terms of the LCWFs in \eq\eqref{eq: quark in quark PDF g expansion NLO} to the definition of the splitting functions that we want to extract.

We start with the definition of a bare PDF for parton $i$ in a hadronic state:
\begin{equation}\label{barepdf}
    f^0_{i/h}(\mu,x)\equiv\frac{1}{4\pi x}\int\frac{\mathrm{d}^{D-2}\mathbf{k}}{(2\pi)^{D-2}}\frac{\left\langle h(p^+)\right|a_i^{\dagger}(xp^+,\mathbf{k})a_{i}(xp^+,\mathbf{k})\left|h(p^+)\right\rangle}{\left\langle h(p^+)|h(p^+)\right\rangle}.
\end{equation}
Here the creation and annihilation operators are bare ones, and the definition is UV divergent, regularized by keeping $\varepsilon> 0$. This also introduces a dependence on the dimensional regularization scale $\mu$, needed to keep the PDF and the coupling constant dimensionless. The general statement of (UV) renormalization of the PDF~\cite{Collins:2011zzd} is that it is possible to find a matrix $Z_{ii'}(\mu_F,\mu,x)$ such that the renormalized PDF
\begin{equation}\label{renormpdf}
    f^R_{i/h}(\mu_F) = Z_{ii'}(\mu^2/\mu_F^2)\otimes f^0_{i'/h}
\end{equation}
is UV finite. In addition to canceling the divergent parts, one can include different finite parts in $Z_{ii'}(\mu^2/\mu_F,x)$, corresponding to different renormalization schemes. The bare PDF \eqref{barepdf} will depend on the regularization scale $\mu$, which is arbitrary. Thus the dependence on $\mu$ should cancel in physical observables together with the $1/\varepsilon$ poles. When we choose $Z_{ii'}(\mu^2/\mu_F^2,x)$ to cancel the dependence on $\mu$, it must for dimensional reasons  also acquire a dependence on a factorization scale $\mu_F^2$. This  corresponds to the scale at which we impose the renormalization condition. It is the dependence of the renormalization factor on the ratio $\mu^2/\mu_F^2$ that relates the UV divergences of the bare distribution to the actual physical content of DGLAP evolution, which is related to the dependence on the factorization scale $\mu_F^2$. This is the solution of the seeming paradox of why \emph{ultraviolet} renormalization is actually related to the  resummation of large \emph{collinear} logarithms in cross sections.

The DGLAP evolution equation describes the dependence of the PDF's on the factorization scale. It can be obtained by differentiating \eq\eqref{renormpdf} and using the fact that the bare PDF $f^0_{i/h}$ does not depend on $\mu_F$:
\begin{equation}\label{dglap}
 \frac{\ud}{\ud \ln \mu_F^2}   f^R_{i/h}(\mu_F) = 
\left(\frac{\ud}{\ud \ln \mu_F^2} Z_{ii'}(\mu^2/\mu_F^2)\right)\otimes f^0_{i'/h}
= \left(\frac{\ud}{\ud \ln \mu_F^2} Z_{ii'}(\mu^2/\mu_F^2)\right)\otimes Z^{-1}_{i'j}(\mu^2/\mu_F^2)  \otimes f^R_{j/h}.
\end{equation}
This enables us to obtain the definition of the DGLAP splitting functions
\begin{equation}\label{defP}
P_{ij} = \left(\frac{\ud}{\ud \ln \mu_F^2} Z_{ii'}(\mu^2/\mu_F^2)\right)\otimes Z^{-1}_{i'j}(\mu^2/\mu_F^2).
\end{equation}
Equations \eqref{renormpdf} and \eqref{defP} provide the general definition of a splitting function. Since they do not depend on the hadronic target $h$, we can evaluate them for any hadronic target, in particular a dressed quark state 
$|q(p^+)\rangle_\mathrm{int}$.

When we use  a dressed parton state that is spacelike off-shell with a virtuality $\mu_0^2$, a  natural condition to fix also the finite parts of the renormalization matrix $Z_{ii'}(\mu^2/\mu_F^2,x)$ is  to demand the renormalization condition
\begin{equation}\label{eq:offshellpole}
 \left. f^R_{i/i(\mu_0)}(\mu_F,x)\right|_{\mu_F=\mu_0} = \delta(1-x).
\end{equation}
This would open up the possibility of using the off-shellness of the incoming parton as a way to directly implement collinear factorization in a cross section calculation without the need for a separate subtraction of collinear divergences.\footnote{An idea suggested to us by G. Beuf in a private communication long ago.} 
However, in this paper we are only concerned with the divergent parts and logarithms, and leave a further discussion of this possibility to future work.

For simplicity, we focus from now on only on the non-singlet (i.e. valence)  distribution $f_{-/h}$. It  decouples from the gluon, and thus for this distribution the renormalization matrix $Z_{ij}$ becomes just a number $Z_V$.  What we actually get from the loop diagrams is the bare distribution in a dressed quark state that is off-shell with an invariant mass $-\mu_0^2$. In general the  bare valence quark (quark minus antiquark) distribution in a dressed quark state is, including the finite parts, of the form 
\begin{multline}\label{eq:fzero}
  f^0_{-/q(\mu_0)}(\mu,x) = \delta(1-x) + \left(\frac{\as}{2\pi}\right)
  \left[\left(\frac{1}{\varepsilon} + \ln \frac{\mu^2}{\mu_0^2}\right)P^{-,(0)}(x) + C^{(0)}_{-/q}(x)\right]
+ \frac{1}{2}\left(\frac{\as}{2\pi}\right)^2 
\left[\left(\frac{1}{\varepsilon} + \ln \frac{\mu^2}{\mu_0^2}
\right)P^{-,(0)} + C^{(0)}_{-/q}\right]^2(x)
\\  
+ \frac{1}{2}\left(\frac{\as}{2\pi}\right)^2 
\left[\left(\frac{1}{\varepsilon} + 2 \ln \frac{\mu^2}{\mu_0^2}\right)P^{-,(1)}(x)
+ C^{(1)}_{-/q}(x)\right]
+ \mathcal{O}(\as^3),
\end{multline}
where the square should be understood as a convolution.
Here the combination $\left(\frac{1}{\varepsilon} + 2 \ln \frac{\mu^2}{\mu_F^2}\right)$ at the two-loop level comes from the observation that dimensionally a two-loop integral has to be proportional to $\mu^{4\varepsilon}$, and thus the single pole in a two-loop calculation will be accompanied by $2 \ln \frac{\mu^2}{\mu_F^2}$. This equation reduces to  \eq\eqref{eq: pdf in terms of splitting functions} when all the scales are taken to be equal, which allows us to read off the splitting functions from the calculation of  $f^0_{-/q}(\mu,x)$ .

For this nonsinglet distribution   renormalization can be achieved with the following choice for the renormalization factor:
\begin{eqnarray}
 Z_{V}(\mu^2/\mu_F^2,x) &=& 
 %\delta(1-x) - \left(\frac{\as}{2\pi}\right)\left(\frac{1}{\varepsilon} + \ln \frac{\mu^2}{\mu_F^2}\right)P^{-,(0)}(x)
%- \frac{1}{2}\left(\frac{\as}{2\pi}\right)^2 
%\left(\frac{1}{\varepsilon} + 2 \ln \frac{\mu^2}{\mu_F^2}\right)P^{-,(1)}(x)
%\\ \nonumber && \quad 
%+ \frac{1}{2}\left(\frac{\as}{2\pi}\right)^2 
%\left(\frac{1}{\varepsilon} + \ln \frac{\mu^2}{\mu_F^2}\right)^2 
%\left(P^{-,(0)}\otimes P^{-,(0)}\right)(x)
%+ \mathcal{O}(\as^3)
\delta(1-x) - \left(\frac{\as}{2\pi}\right)
  \left[\left(\frac{1}{\varepsilon} + \ln \frac{\mu^2}{\mu_F^2}\right)P^{-,(0)}(x) + C^{(0)}_{-/q}(x)\right]
+ \frac{1}{2}\left(\frac{\as}{2\pi}\right)^2 
\left[\left(\frac{1}{\varepsilon} + \ln \frac{\mu^2}{\mu_F^2}
\right)P^{-,(0)} + C^{(0)}_{-/q}\right]^2(x)
\nonumber \\   && \quad 
- \frac{1}{2}\left(\frac{\as}{2\pi}\right)^2 
\left[\left(\frac{1}{\varepsilon} + 2 \ln \frac{\mu^2}{\mu_F^2}\right)P^{-,(1)}(x)
+ C^{(1)}_{-/q}(x)\right]
+ \mathcal{O}(\as^3),
\\
\label{eq:zminusone}
 Z^{-1}_{V}(\mu^2/\mu_F^2,x) &=& 
\delta(1-x) + \left(\frac{\as}{2\pi}\right)
  \left[\left(\frac{1}{\varepsilon} + \ln \frac{\mu^2}{\mu_F^2}\right)P^{-,(0)}(x) + C^{(0)}_{-/q}(x)\right]
+ \frac{1}{2}\left(\frac{\as}{2\pi}\right)^2 
\left[\left(\frac{1}{\varepsilon} + \ln \frac{\mu^2}{\mu_F^2}
\right)P^{-,(0)} + C^{(0)}_{-/q}\right]^2(x)
\nonumber \\    && \quad 
+ \frac{1}{2}\left(\frac{\as}{2\pi}\right)^2 
\left[\left(\frac{1}{\varepsilon} + 2 \ln \frac{\mu^2}{\mu_F^2}\right)P^{-,(1)}(x)
+ C^{(1)}_{-/q}(x)\right]
+ \mathcal{O}(\as^3).
 %\delta(1-x) + \left(\frac{\as}{2\pi}\right)\left(\frac{1}{\varepsilon} + \ln \frac{\mu^2}{\mu_F^2}\right)P^{-,(0)}(x)
%+ \frac{1}{2}\left(\frac{\as}{2\pi}\right)^2 
%\left(\frac{1}{\varepsilon} + 2 \ln \frac{\mu^2}{\mu_F^2}\right)P^{-,(1)}(x)
%\\ \nonumber  && \quad 
%+ \frac{1}{2}\left(\frac{\as}{2\pi}\right)^2 
%\left(\frac{1}{\varepsilon} + \ln \frac{\mu^2}{\mu_F^2}\right)^2 
%\left(P^{-,(0)}\otimes P^{-,(0)}\right)(x)
%+ \mathcal{O}(\as^3).
\end{eqnarray}
It is easy to see that these equations  lead both to the renormalization condition  \eqref{eq:offshellpole}, and to the DGLAP equation with the desired normalization 
\begin{equation}
    P^{-}(x) = \left(\frac{\ud}{\ud \ln \mu_F^2} Z_{V}(\mu^2/\mu_F^2)\right)\otimes Z^{-1}_{V}(\mu^2/\mu_F^2) 
    = \left(\frac{\as}{2\pi}\right) P^{-,(0)}(x) +  \left(\frac{\as}{2\pi}\right)^2 P^{-,(1)}(x) + \mathcal{O}(\as^3).
\end{equation}
 Note that a naive interpretation of DGLAP evolution as describing the dependence on the regularization scale $\mu$ rather than the correct understanding in terms of the factorization scale $\mu_F$ gives the wrong sign. 
It would also be tempting to just differentiate $Z^{-1}_{V}$ with respect to $\mu^2$, rather than $Z_{V}$ with respect to $\mu_F^2$. This will indeed lead to the right coefficients $P^{-,(0)}$ and $P^{-,(1)}$ and thus allow one to read off the correct splitting functions from a calculation of the LCWF at this order. But this way of thinking will garble the interpretation of the sign of the $P^{-,(0)} \otimes P^{-,(0)}$ term.

The essential lesson from these manipulations is that the actual definition of the splitting function~\eqref{defP} is in terms of derivatives of the renormalization factor $Z_{ii'}$, rather than a derivative of the bare distribution in \eq\eqref{eq: pdf in terms of splitting functions}, or \eqref{eq:fzero}. The $\as^2/\varepsilon$ term of the latter also includes a convolution 
\begin{equation}
\left(\frac{\as}{2\pi}\right)^2 
\frac{1}{\varepsilon}P^{-,(0)} \otimes C^{(0)}_{-/q}(x)
\end{equation}
between the leading order splitting function and the finite part of the  leading order correction to the bare distribution, which must be removed by the matching counterterm in \eq\eqref{eq: matching counterterm} in order to access the NLO splitting function $P^{-,(1)}$. 

Let us finally allow ourselves a small digression into these finite parts $C^{(0)}_{-/q}(x), C^{(1)}_{-/q}(x), \dots$ in the renormalization factor. 
Firstly, in the case of on-shell incoming partons and massless quarks, this ``finite'' renormalization is in fact, confusingly, collinearly divergent. To make things more complicated, in dimensional regularization this collinear IR pole  tends to cancel the UV divergence that one wants to extract and renormalize, unless special care is taken in the calculation. This tends to make presentations in the literature of PDF renormalization in terms of the operator definition  rather baroque, involving various kinds of mysterious, seemingly ad hoc ``counterterms'' or ``subtraction terms''. As a result many authors prefer to dig out the splitting function from collinear divergences in a physical scattering process in stead (see e.g.~\cite{Paukkunen:2009ks}). Secondly, one could also wonder whether the appearance of the matching counterterm is an artefact of choosing the natural renormalization condition \eqref{eq:offshellpole},  which leads to the renormalization factor also including a finite part $C^{(0)}$. 
It might first  seem so since the finite part appears in \eq\eqref{eq: pdf in terms of splitting functions}, or \eqref{eq:fzero}, from which we read off the renormalization factor in terms of the LCWFs. However, the finite part in fact cancels out, at least at this order, when using \eq\eqref{defP} to calculate the splitting function from the renormalization factor $Z_{ii'}$.
It would also be possible to choose a different renormalization condition, choosing for example the finite term in $Z_{ii'}$ to be zero. In this case we should first express the delta function $\delta(1-x)$ in terms of the renormalized PDF, and substitute this expression  on the r.h.s. of 
\eq\eqref{eq: pdf in terms of splitting functions}, or \eqref{eq:fzero}. Only this would transform 
\eq\eqref{eq: pdf in terms of splitting functions},  or \eqref{eq:fzero}, into the form 
$f^0_{i/h} = Z^{-1}_{ii'} f_{i'/h}$ and enable us to read off $Z^{-1}_{ii'}$  from the calculation of $f^0_{i/h}$ and determine $Z_{ii'}$. In the end this would lead to the same form for the bare quark distribution in a dressed quark $f^0_{i/q(\mu_0)}$ in terms of the splitting functions. However, the interpretation of \eq\eqref{eq: pdf in terms of splitting functions}, or \eqref{eq:fzero}, would be different since the $\delta(1-x)$ appearing there would not any more be the renormalized quark distribution at $\mu_F=\mu_0$.
Both approaches would require us to add the same matching term, \eq\eqref{eq: matching counterterm}, to the expression obtained from squaring the LCWFs in order to extract the splitting function. Ultimately the square of the LO term, including its finite part, unavoidably appears in our expressions because the splitting function~\eqref{defP} is a nonlinear function of the renormalization factor $Z$.

%%%%%%%%%%%%%%%%%%%%%%%%%%%%%%%%%%%%%%%%%%%%%%

\section{Two-loop integrals in LCPT}\label{Appendix: integrals}

In this Appendix we explain how the transverse momentum loop integrals in this paper are evaluated. The tensor integrals are reduced to scalar integrals with the Passarino-Veltman method \cite{Passarino:1978jh}, and the remaining scalar integrals are evaluated by relating them to a master integral. Although the procedure is very similar to familiar loop integrations in covariant theory, we present here the key steps in detail, due to there being two major differences between LCPT and covariant theory integrals: First, the tensor reduction is based on rotational symmetry in $D-2$ dimensions instead of the full Lorentz symmetry. Second, the denominator structures in the remaining scalar integrals are more complicated, since the LCPT energy denominators ``remember'' preceding intermediate states of the process, unlike ordinary propagators.

As described in section \ref{Section: Detailed example}, the general two loop integrals we encounter are of the form 
    \begin{equation}
    \mathcal{I}=\left(\bar\mu^{4-D}\right)^2\int \frac{\mathrm{d}^{D-2}\mathbf{l}}{(2\pi)^{D-2}}\frac{1}{\mathbf{l}^2+\Delta_0\mu^2} \int\frac{\mathrm{d}^{D-2}\mathbf{k}}{(2\pi)^{D-2}}\frac{N_1\mathbf{k}^2\mathbf{l}^2+N_2\mathbf{k}^2(\mathbf{k}\cdot\mathbf{l})+N_3(\mathbf{k}\cdot\mathbf{l})\mathbf{l}^2+N_4(\mathbf{k}\cdot\mathbf{l})^2}{\mathbf{k}^2\left[\mathbf{k}^2+\Delta_1(\mathbf{k}+\Delta_2\mathbf{l})^2\right]\left[(\mathbf{k}+\Delta_3\mathbf{l})^2+\Delta_4\mathbf{l}^2\right]}.
    \end{equation}
Thus, for any given loop in a diagram, the integral has at most three denominators that depend on the loop momentum, and the rank of the tensor integral is at most two. We introduce the following notation for the integrals:
\begin{align}
    \mathcal{A}_0(\Delta_1,\Delta_{2},\mathbf{l})&\equiv\bar\mu^{4-D}\int\frac{\mathrm{d}^{D-2}\mathbf{k}}{(2\pi)^{D-2}}\frac{1}{\left[ \mathbf{k}^2+\Delta_1\left(\mathbf{k}+\Delta_2\mathbf{l}\right)^2\right]},\\
    %%%
    \mathcal{B}_0(\Delta_1,\Delta_{2},\Delta_3,\Delta_4,\mathbf{l})&\equiv\bar\mu^{4-D}\int\frac{\mathrm{d}^{D-2}\mathbf{k}}{(2\pi)^{D-2}}\frac{1}{\left[ \mathbf{k}^2+\Delta_1\left(\mathbf{k}+\Delta_2\mathbf{l}\right)^2\right]\left[\left(\mathbf{k}+\Delta_3\mathbf{l}\right)^2+\Delta_4\mathbf{l}^2\right]},\\
    %%%
    \mathcal{B}^{i}(\Delta_1,\Delta_{2},\Delta_3,\Delta_4,\mathbf{l})&\equiv\bar\mu^{4-D}\int\frac{\mathrm{d}^{D-2}\mathbf{k}}{(2\pi)^{D-2}}\frac{k^i}{\left[ \mathbf{k}^2+\Delta_1\left(\mathbf{k}+\Delta_2\mathbf{l}\right)^2\right]\left[\left(\mathbf{k}+\Delta_3\mathbf{l}\right)^2+\Delta_4\mathbf{l}^2\right]},\\
    %%%
    \mathcal{B}^{ij}(\Delta_1,\Delta_{2},\Delta_3,\Delta_4,\mathbf{l})&\equiv\bar\mu^{4-D}\int\frac{\mathrm{d}^{D-2}\mathbf{k}}{(2\pi)^{D-2}}\frac{k^ik^j}{\left[ \mathbf{k}^2+\Delta_1\left(\mathbf{k}+\Delta_2\mathbf{l}\right)^2\right]\left[\left(\mathbf{k}+\Delta_3\mathbf{l}\right)^2+\Delta_4\mathbf{l}^2\right]},\label{Eq: Bij}\\
    %%%
    \mathcal{C}_0(\Delta_1,\Delta_{2},\Delta_3,\Delta_4,\mathbf{l})&\equiv\bar\mu^{4-D}\int\frac{\mathrm{d}^{D-2}\mathbf{k}}{(2\pi)^{D-2}}\frac{1}{\mathbf{k}^2\left[ \mathbf{k}^2+\Delta_1\left(\mathbf{k}+\Delta_2\mathbf{l}\right)^2\right]\left[\left(\mathbf{k}+\Delta_3\mathbf{l}\right)^2+\Delta_4\mathbf{l}^2\right]},\label{Eq: C0}\\
    %%%
    \mathcal{C}^{i}(\Delta_1,\Delta_{2},\Delta_3,\Delta_4,\mathbf{l})&\equiv\bar\mu^{4-D}\int\frac{\mathrm{d}^{D-2}\mathbf{k}}{(2\pi)^{D-2}}\frac{k^i}{\mathbf{k}^2\left[ \mathbf{k}^2+\Delta_1\left(\mathbf{k}+\Delta_2\mathbf{l}\right)^2\right]\left[\left(\mathbf{k}+\Delta_3\mathbf{l}\right)^2+\Delta_4\mathbf{l}^2\right]},\\
    %%%
    \mathcal{C}^{ij}(\Delta_1,\Delta_{2},\Delta_3,\Delta_4,\mathbf{l})&\equiv\bar\mu^{4-D}\int\frac{\mathrm{d}^{D-2}\mathbf{k}}{(2\pi)^{D-2}}\frac{k^ik^j}{\mathbf{k}^2\left[ \mathbf{k}^2+\Delta_1\left(\mathbf{k}+\Delta_2\mathbf{l}\right)^2\right]\left[\left(\mathbf{k}+\Delta_3\mathbf{l}\right)^2+\Delta_4\mathbf{l}^2\right]}.
\end{align}

\subsection{Tensor reduction}

Let us consider the reduction of the most complicated tensor integral $\mathcal{C}^{ij}$. Due to rotational invariance in the transverse momentum plane, the tensor structure must be
\begin{equation}\label{eq: C tensor structure}
    \mathcal{C}^{ij}(\Delta_1,\Delta_{2},\Delta_3,\Delta_4,\mathbf{l})=l^il^j\mathcal{C}_{21}(\Delta_1,\Delta_{2},\Delta_3,\Delta_4,\mathbf{l})+\delta^{ij}\mathcal{C}_{22}(\Delta_1,\Delta_{2},\Delta_3,\Delta_4,\mathbf{l}),
\end{equation}
where $\mathcal{C}_{21}$ and $\mathcal{C}_{22}$ are scalar integrals to be determined, for which we need two equations. The first one is obtained by multiplying Eq.~\eqref{eq: C tensor structure} by $\delta^{ij}$, giving
\begin{equation}\label{Eq: C integral first eq}
    \mathbf{l}^2\mathcal{C}_{21}+(D-2)\mathcal{C}_{22}=\bar\mu^{4-D}\int\frac{\mathrm{d}^{D-2}\mathbf{k}}{(2\pi)^{D-2}}\frac{1}{\left[ \mathbf{k}^2+\Delta_1\left(\mathbf{k}+\Delta_2\mathbf{l}\right)^2\right]\left[\left(\mathbf{k}+\Delta_3\mathbf{l}\right)^2+\Delta_4\mathbf{l}^2\right]}=\mathcal{B}_0.
\end{equation}
The second equation is obtained by multiplying Eq.~\eqref{eq: C tensor structure} by $l^j$:
\begin{equation}\label{Eq: C integral dot product}
\begin{aligned}
    \mathcal{C}_{21}\mathbf{l}^2l^i+\mathcal{C}_{22}l^i&=\bar\mu^{4-D}\int\frac{\mathrm{d}^{D-2}\mathbf{k}}{(2\pi)^{D-2}}\frac{k^i\mathbf{k}\cdot\mathbf{l} }{\mathbf{k}^2\left[ \mathbf{k}^2+\Delta_1\left(\mathbf{k}+\Delta_2\mathbf{l}\right)^2\right]\left[\left(\mathbf{k}+\Delta_3\mathbf{l}\right)^2+\Delta_4\mathbf{l}^2\right]}\\
    &=\frac{1}{2\Delta_3}\left\{ \mathcal{B}^i(\Delta_1,\Delta_2,0,0,\mathbf{l})-\mathcal{B}^i(\Delta_1,\Delta_2,\Delta_3,\Delta_4,\mathbf{l})-(\Delta_3^2+\Delta_{4})\mathbf{l}^2\mathcal{C}^i(\Delta_1,\Delta_2,\Delta_3,\Delta_4,\mathbf{l})\right\},
\end{aligned}
\end{equation}
where we substituted in the dot product
\begin{equation}\label{eq: reduction l dot k}
    \mathbf{k}\cdot\mathbf{l}=\frac{1}{2\Delta_3}\left\{\left[(\mathbf{k}+\Delta_3\mathbf{l})^2+\Delta_4\mathbf{l}^2 \right] -\left[\mathbf{k}^2\right]-\left[\left(\Delta_3^2+\Delta_4\right)\mathbf{l}^2\right]   \right\},
\end{equation}
which is solved from the square
\begin{equation}
    (\mathbf{k}+\Delta_3\mathbf{l})^2=\mathbf{k}^2+\Delta_3^2\mathbf{l}^2+2\Delta_3\mathbf{k}\cdot\mathbf{l}.
\end{equation}
On the other hand, the tensor structure for the rank one integrals must be
\begin{align}
    \mathcal{B}^{i}(\Delta_1,\Delta_{2},\Delta_3,\Delta_4,\mathbf{l})&=l^i\mathcal{B}_1(\Delta_1,\Delta_{2},\Delta_3,\Delta_4,\mathbf{l}),\label{Eq: Bi tensor structure}\\
    \mathcal{C}^{i}(\Delta_1,\Delta_{2},\Delta_3,\Delta_4,\mathbf{l})&=l^i\mathcal{C}_1(\Delta_1,\Delta_{2},\Delta_3,\Delta_4,\mathbf{l}),\label{Eq: Ci tensor structure} 
\end{align}
so that Eq. \eqref{Eq: C integral dot product} becomes
\begin{equation}\label{Eq: C integral second eq}
    \mathcal{C}_{21}\mathbf{l}^2+\mathcal{C}_{22}=\frac{1}{2\Delta_3}\left\{ \mathcal{B}_1(\Delta_1,\Delta_2,0,0,\mathbf{l})-\mathcal{B}_1(\Delta_1,\Delta_2,\Delta_3,\Delta_4,\mathbf{l})-(\Delta_3^2+\Delta_{4})\mathbf{l}^2\mathcal{C}_1(\Delta_1,\Delta_2,\Delta_3,\Delta_4,\mathbf{l})\right\}.
\end{equation}
Now, solving the system of equations \eqref{Eq: C integral first eq} and \eqref{Eq: C integral second eq} gives the expressions for the scalar integrals:
\begin{align}
    \mathcal{C}_{21}&=-\frac{1}{2(D-3)\Delta_3\mathbf{l}^2}\Big\{(D-2)\big[\mathcal{B}_1(\Delta_1,\Delta_2,\Delta_3,\Delta_4,\mathbf{l})-\mathcal{B}_1(\Delta_1,\Delta_2,0,0,\mathbf{l})+(\Delta_3^2+\Delta_{4})\mathbf{l}^2\mathcal{C}_1(\Delta_1,\Delta_2,\Delta_3,\Delta_4,\mathbf{l})\big]\nonumber\\
    &\qquad\qquad\qquad\qquad\quad+2\Delta_3\mathcal{B}_0(\Delta_1,\Delta_2,\Delta_3,\Delta_4,\mathbf{l})\Big\},\\
    \mathcal{C}_{22}&=\frac{1}{2(D-3)\Delta_3}\Big\{\big[\mathcal{B}_1(\Delta_1,\Delta_2,\Delta_3,\Delta_4,\mathbf{l})-\mathcal{B}_1(\Delta_1,\Delta_2,0,0,\mathbf{l})+(\Delta_3^2+\Delta_{4})\mathbf{l}^2\mathcal{C}_1(\Delta_1,\Delta_2,\Delta_3,\Delta_4,\mathbf{l})\big]\nonumber\\
    &\qquad\qquad\qquad\quad+2\Delta_3\mathcal{B}_0(\Delta_1,\Delta_2,\Delta_3,\Delta_4,\mathbf{l})\Big\}.
\end{align}
The remaining scalar integrals $\mathcal{B}_1$ and $\mathcal{C}_1$ can be derived in a similar way by multiplying Eqs. \eqref{Eq: Bi tensor structure} and \eqref{Eq: Ci tensor structure} by $l^j$ and substituting in the dot product \eqref{eq: reduction l dot k}. The result is
\begin{align}
    \mathcal{B}_1(\Delta_1,\Delta_2,\Delta_3,\Delta_4,\mathbf{l})&=\frac{1}{\Delta_3\mathbf{l}^2}\left\{\mathcal{A}_0(\Delta_1,\Delta_2,\mathbf{l})-B^{ii}(\Delta_1,\Delta_2,\Delta_3,\Delta_4,\mathbf{l})-\left(\Delta_3^2+\Delta_4\right)\mathbf{l}^2\mathcal{B}_0(\Delta_1,\Delta_2,\Delta_3,\Delta_4,\mathbf{l})\right\},\\
    \mathcal{C}_1(\Delta_1,\Delta_2,\Delta_3,\Delta_4,\mathbf{l})&=\frac{1}{\Delta_3\mathbf{l}^2}\left\{\mathcal{B}_0(\Delta_1,\Delta_2,0,0,\mathbf{l})-B_0(\Delta_1,\Delta_2,\Delta_3,\Delta_4,\mathbf{l})-\left(\Delta_3^2+\Delta_4\right)\mathbf{l}^2\mathcal{C}_0(\Delta_1,\Delta_2,\Delta_3,\Delta_4,\mathbf{l})\right\}.
\end{align}
As a curious aside, here $B^{ii}$ is the rank two tensor integral \eqref{Eq: Bij} with the numerator momenta contracted, making it a scalar integral. Unlike in covariant theory, this integral is not easily expressed in terms of other scalar integrals, but it is, however, straightforward to just evaluate, which is why we have kept it as is in the expression above.

\subsection{Evaluation of scalar integrals}
Let us then briefly demonstrate how the remaining scalar integrals are evaluated. To this end, we define a master integral formula
\begin{equation}\label{Eq: master integral}
    \Phi(a,b,\Delta)\equiv\mu^{4-D}\int\frac{\text{d}^{D-2}\mathbf{l}}{(2\pi)^{D-2}}\frac{\mathbf{l}^{2a}}{(\mathbf{l}^2+\Delta)^b}=\frac{\mu^{4-D}}{(4\pi)^{\frac{D}{2}-1}}\frac{1}{\Delta^{b-a-\frac{D}{2}+1}}\frac{\Gamma\left(a+\frac{D}{2}-1\right)\Gamma\left(b-a-\frac{D}{2}+1\right)}{\Gamma(b)\Gamma(\frac{D}{2}-1)},
\end{equation}
which is a standard field theory integral. After Feynman parametrization, each scalar integral can be written in terms of the master integral after a suitable change of variables. Furthermore, in the two-loop diagrams evaluated in this paper, after performing the first loop integration, the second loop integral is always of the form of this master integral, with either $a$ or $b$ depending on $\varepsilon$.

As an example, let us again consider the most complicated scalar integral we need to evaluate, $\mathcal{C}_0$, given by
\begin{equation}
        \mathcal{C}_0(\Delta_1,\Delta_{2},\Delta_3,\Delta_4,\mathbf{k})\equiv\bar\mu^{4-D}\int\frac{\mathrm{d}^{D-2}\mathbf{l}}{(2\pi)^{D-2}}\frac{1}{\mathbf{l}^2\left[ \mathbf{l}^2+\Delta_1\left(\mathbf{l}+\Delta_2\mathbf{k}\right)^2\right]\left[\left(\mathbf{l}+\Delta_3\mathbf{k}\right)^2+\Delta_4\mathbf{k}^2\right]}.\\
\end{equation}
To evaluate this, we introduce two Feynman parameters $y_1$ and $y_2$, complete the square in the denominator, and shift the integration variable as
\begin{equation}
    \mathbf{l}\rightarrow\tilde{\mathbf{l}}=\mathbf{l}+\left[y_1\Delta_3+(1+y_2)\frac{\Delta_1\Delta_2}{1+\Delta_1}\right]\mathbf{k},
\end{equation}
which after some algebra gives
\begin{equation}
\begin{aligned}
    \mathcal{C}_0(\Delta_1,\Delta_{2},\Delta_3,\Delta_4,\mathbf{k})&=\frac{2}{1+\Delta_1}\int_0^1\mathrm{d}y_1\int_0^{1-y_1}\mathrm{d}y_2\int\frac{\mathrm{d}^{D-2}\tilde{\mathbf{l}}}{(2\pi)^{D-2}}\frac{1}{\left[\tilde{\mathbf{l}}^2+\tilde{\Delta}\mathbf{k}^2\right]^3}\\
    &=\frac{2}{1+\Delta_1}\int_0^1\mathrm{d}y_1\int_0^{1-y_1}\mathrm{d}y_2\Phi(0,3,\tilde{\Delta}\mathbf{k}^2),
\end{aligned}
\end{equation}
where
\begin{equation}
    \tilde{\Delta}=\frac{1}{(1+\Delta_1)^2}\left[(1-y_2)y_2\Delta_1^2\Delta_2^2+y_2\Delta_1\Delta_2(\Delta_2-2y_1(1+\Delta_1)\Delta_3)   \right]+y_1(1-y_1)\Delta_3^2+y_1\Delta_4.
\end{equation}

All the other scalar integrals can be evaluated in a similar fashion. Although the integrals are in principle simple, the energy denominators introduce weight factors $\Delta_i$ between different momenta, which result in rather complicated expressions for the final elementary integrals. As a technical challenge to the automatization procedure, the remaining Feynman parameter integrals (performed while keeping $D=4-2\varepsilon$) result in hypergeometric functions, which can be difficult to evaluate efficiently.

\section{Detailed results for diagram calculations}\label{appendix: diagram results}
In this Appendix we present results for each diagram shown in Fig.~\ref{fig: all diagrams}. For each diagram, we give an expression that makes the numerator and integral structure explicit and serves as the input to the evaluation process outlined in section \ref{Section: Detailed example}. We then show the resulting contribution to $f^-$.

First, we give the contributions from the virtual one-loop corrections to the $qg$ wave function, diagrams A1--A7. The quark loop correction to the gluon propagator, diagram A1, gives
\begin{equation}
\begin{aligned}
    &f^-_\text{A1}\Big|_{1/\varepsilon}=2\as ^2 C_FN_fT_Fx^2\int_0^{1-x}\mathrm{d}\xi
    \left(\bar{\mu}^{4-D}\right)^2\int \frac{\mathrm{d}^{D-2}\mathbf{k}}{(2\pi)^{D-2}}\frac{1}{\left[\mathbf{k}^2+x(1-x)\mu^2\right]^3}\\
     &\times\int\frac{\mathrm{d}^{D-2}\mathbf{l}}{(2\pi)^{D-2}}\frac{[\bar{u}_{h}(p)\slashed\varepsilon_{\lambda}(q)u_{h'}(k)][\bar{v}_{s'}(l')\slashed\varepsilon_{\lambda}^*(q)u_{s}(l)][\bar{u}_{s}(l)\slashed\varepsilon_{\lambda'}(q)v_{s'}(l')][\bar{u}_{h'}(k)\slashed\varepsilon_{\lambda'}^*(q)u_h(p)]}{\left[\left(\mathbf{l}+\frac{\xi}{1-x}\mathbf{k}\right)^2+\frac{\xi(1-x-\xi)}{x(1-x)^2}\left(\mathbf{k}^2+x(1-x)\mu^2\right)\right]}\Big|_{1/\varepsilon}\\
     &=\frac{1}{2\varepsilon}\frac{\as ^2}{(2\pi)^2}C_FN_fT_F\bigg\{\frac{1+x^2}{1-x}\left[\frac{4}{9}+\frac{4}{3}\ln x+\frac{8}{3}\ln (1-x)\right]+\frac{4}{3}(1-x)\bigg\}.
\end{aligned}
\end{equation}

The gluon loop correction to the gluon propagator, diagram A2, gives
\begin{equation}
\begin{aligned}
    &f^-_\text{A2}\Big|_{1/\varepsilon}=4\as ^2 C_FC_Ax^2\int_\alpha^{1-x-\alpha}\mathrm{d}\xi
    \left(\bar{\mu}^{4-D}\right)^2\int \frac{\mathrm{d}^{D-2}\mathbf{k}}{(2\pi)^{D-2}}\frac{1}{\left[\mathbf{k}^2+x(1-x)\mu^2\right]^3}\\
     &\times\int\frac{\mathrm{d}^{D-2}\mathbf{l}}{(2\pi)^{D-2}}\frac{[\bar{u}_{h}(p)\slashed\varepsilon_{\lambda}(q)u_{h'}(k)][\bar{u}_{h'}(k)\slashed\varepsilon_{\lambda'}^*(q)u_h(p)]\Gamma_{\lambda';\sigma,\sigma'}(\vec{q};\vec{l},\vec{l}\,')\Gamma^\dagger_{\lambda;\sigma,\sigma'}(\vec{q};\vec{l},\vec{l}\,')}{\left[\left(\mathbf{l}+\frac{\xi}{1-x}\mathbf{k}\right)^2+\frac{\xi(1-x-\xi)}{x(1-x)^2}\left(\mathbf{k}^2+x(1-x)\mu^2\right)\right]}\Big|_{1/\varepsilon}\\
     &=\frac{1}{2\varepsilon}\frac{\as ^2}{(2\pi)^2}C_FC_A\bigg\{\frac{1+x^2}{1-x}\bigg[\frac{1}{9}-\frac{2\pi^2}{3}-\frac{11}{3}\ln(x)+\frac{2}{3}\ln(1-x)+4\ln(x)\ln(1-x)+6\ln^2(1-x)\\
     &-8\ln\alpha-4\ln(\alpha)\ln(x(1-x))-2\ln^2(\alpha) \bigg]+(1-x)\left[-\frac{11}{3}+4\ln\left(\frac{1-x}{\alpha}\right)\right]\bigg\},
\end{aligned}
\end{equation}
where we introduced a shorthand notation for the $1\rightarrow 2$ gluon splitting vertex 
\begin{equation}
    \Gamma_{\lambda_1;\lambda_2,\lambda_3}(\vec{p}_1;\vec{p}_2,\vec{p}_3) \equiv\left[\frac{\varepsilon_{\lambda_2}^{*j}\varepsilon_{\lambda_3}^{*k}\varepsilon_{\lambda_1}^{l}}{1-p_2^+/p_1^+}+\frac{\varepsilon_{\lambda_3}^{*j}\varepsilon_{\lambda_2}^{*k}\varepsilon_{\lambda_1}^{l}}{p_2^+/p_1^+}-\varepsilon_{\lambda_1}^{j}\varepsilon_{\lambda_3}^{*k}\varepsilon_{\lambda_2}^{*l}\right]\delta^{ij}\delta^{kl}\left(p_2^i-\frac{p_2^+}{p_1^+}p_1^i\right),
\end{equation}
which omits the coefficient with the coupling and color factor, so that the full vertex is $-2igf^{abc}\Gamma_{\lambda_1;\lambda_2,\lambda_3}(\vec{p}_1;\vec{p}_2,\vec{p}_3)$.

The correction to the measured quark propagator, diagram A3, is
\begin{equation}
\begin{aligned}
    &f^-_\text{A3}\Big|_{1/\varepsilon}=2\as ^2 C_F^2(1-x)^2\int_\alpha^{x-\alpha}\mathrm{d}\xi
    \left(\bar{\mu}^{4-D}\right)^2\int \frac{\mathrm{d}^{D-2}\mathbf{k}}{(2\pi)^{D-2}}\frac{1}{\left[\mathbf{k}^2+x(1-x)\mu^2\right]^3}\\
     &\times\int\frac{\mathrm{d}^{D-2}\mathbf{l}}{(2\pi)^{D-2}}\frac{[\bar{u}_{h}(p)\slashed\varepsilon_{\lambda}(q)u_{h'}(k)][\bar{u}_{h'}(k)\slashed\varepsilon_{\lambda'}(l)u_{s'}(l')][\bar{u}_{s'}(l')\slashed\varepsilon^*_{\lambda'}(l)u_{s}(k)][\bar{u}_{s}(k)\slashed\varepsilon_{\lambda}^*(q)u_h(p)]}{\left[\left(\mathbf{l}-\frac{\xi}{x}\mathbf{k}\right)^2+\frac{\xi(x-\xi)}{x^2(1-x)}\left(\mathbf{k}^2+x(1-x)\mu^2\right)\right]}\Big|_{1/\varepsilon}\\
     &=\frac{1}{2\varepsilon}\frac{\as ^2}{(2\pi)^2}C_F^2\bigg\{\frac{1+x^2}{1-x}\Big[1-\frac{2\pi^2}{3}+2\ln x-3\ln(1-x)+4\ln x\ln(1-x)+6\ln^2x\\
     &-8\ln\alpha-4\ln\alpha\ln(x(1-x))-2\ln^2\alpha\Big]+(1-x)\left[-3+4\ln\left(\frac{x}{\alpha}\right)\right]\bigg\}.
\end{aligned}
\end{equation}

Next, we consider the vertex correction diagrams. The contribution from diagram A4 is 
\begin{equation}
\begin{aligned}
    &f^-_\text{A4}\Big|_{1/\varepsilon}=2\as ^2 C_F\left(C_F-\frac{C_A}{2}\right)(1-x)\int_\alpha^{x-\alpha}\mathrm{d}\xi \xi
    \left(\bar{\mu}^{4-D}\right)^2\int \frac{\mathrm{d}^{D-2}\mathbf{k}}{(2\pi)^{D-2}}\frac{1}{\left[\mathbf{k}^2+x(1-x)\mu^2\right]^2}\\
     &\times\int\frac{\mathrm{d}^{D-2}\mathbf{l}}{(2\pi)^{D-2}}\frac{[\bar{u}_{h}(p)\slashed\varepsilon_{\lambda}(q)u_{h'}(k)][\bar{u}_{h'}(k)\slashed\varepsilon_{\lambda'}(l)u_{s'}(l'')][\bar{u}_{s'}(l'')\slashed\varepsilon^*_{\lambda}(q)u_{s}(l')][\bar{u}_{s}(l')\slashed\varepsilon_{\lambda'}^*(l)u_h(p)]}{\mathbf{l}^2\left[\left(\mathbf{l}-\frac{\xi}{x}\mathbf{k}\right)^2+\frac{\xi(x-\xi)}{x^2(1-x)}\left(\mathbf{k}^2+x(1-x)\mu^2\right)\right]}\Big|_{1/\varepsilon}\\
     &=\frac{1}{2\varepsilon}\frac{\as ^2}{(2\pi)^2} C_F\left(C_F-\frac{C_A}{2}\right)\bigg\{\frac{1+x^2}{1-x}\Big[-1+\frac{2\pi^2}{3}-2\ln x+3\ln(1-x)-6\ln x\ln(1-x)-6\ln^2x-2\mathrm{Li}_2(1-x)\\
     &+8\ln\alpha+4\ln\alpha\ln(x(1-x))+2\ln^2\alpha\Big]+1+2x-(10-6x)\ln x-(6-2x)\ln(1-x)+4(1-x)\ln\alpha\bigg\}.
\end{aligned}
\end{equation}

The contribution from diagram A5 is
\begin{equation}
\begin{aligned}
    &f^-_\text{A5}\Big|_{1/\varepsilon}=-2\as ^2C_F\left(C_F-\frac{C_A}{2}\right)x\int_{x}^{1}\mathrm{d}\xi (1-\xi)
    \left(\bar{\mu}^{4-D}\right)^2\int \frac{\mathrm{d}^{D-2}\mathbf{k}}{(2\pi)^{D-2}}\frac{1}{\left[\mathbf{k}^2+x(1-x)\mu^2\right]^2}\\
     &\times\int\frac{\mathrm{d}^{D-2}\mathbf{l}}{(2\pi)^{D-2}}\frac{[\bar{u}_{h}(p)\slashed\varepsilon_{\lambda}(q)u_{h'}(k)][\bar{u}_{h'}(k)\slashed\varepsilon_{\lambda'}(l)v_{s'}(l'')][\bar{v}_{s'}(l'')\slashed\varepsilon^*_{\lambda}(q)u_{s}(l')][\bar{u}_{s}(l')\slashed\varepsilon_{\lambda'}^*(l)u_h(p)]}{\mathbf{l}^2\left[\left(\mathbf{l}-\frac{1-\xi}{1-x}\mathbf{k}\right)^2+\frac{(1-\xi)(\xi-x)}{x(1-x)^2}\left(\mathbf{k}^2+x(1-x)\mu^2\right)\right]}\Big|_{1/\varepsilon}\\
     &=\frac{1}{2\varepsilon}\frac{\as ^2}{(2\pi)^2} C_F\left(C_F-\frac{C_A}{2}\right)\bigg\{\frac{1+x^2}{1-x}\Big[\ln x+4\ln x\ln(1-x)+3\ln^2x\Big]+1-5x+(8-4x)\ln x+(6-2x)\ln(1-x)\bigg\}.
\end{aligned}
\end{equation}

The contribution from diagram A6 is
\begin{equation}
\begin{aligned}
    &f^-_\text{A6}\Big|_{1/\varepsilon}=-2\as ^2C_FC_Ax\int_{\alpha}^{1-x-\alpha}\mathrm{d}\xi\xi
    \left(\bar{\mu}^{4-D}\right)^2\int \frac{\mathrm{d}^{D-2}\mathbf{k}}{(2\pi)^{D-2}}\frac{1}{\left[\mathbf{k}^2+x(1-x)\mu^2\right]^2}\\
     &\times\int\frac{\mathrm{d}^{D-2}\mathbf{l}}{(2\pi)^{D-2}}\frac{[\bar{u}_{h}(p)\slashed\varepsilon_{\lambda}(q)u_{h'}(k)][\bar{u}_{h'}(k)\slashed\varepsilon^*_{\sigma'}(l')u_{s}(l'')][\bar{u}_{s}(l'')\slashed\varepsilon^*_{\sigma}(l)u_{h}(p)]\Gamma^\dagger_{\lambda;\sigma,\sigma'}(\vec{q};\vec{l},\vec{l\,'})}{\mathbf{l}^2\left[\left(\mathbf{l}+\frac{\xi}{1-x}\mathbf{k}\right)^2+\frac{\xi(1-x-\xi)}{x(1-x)^2}\left(\mathbf{k}^2+x(1-x)\mu^2\right)\right]}\Big|_{1/\varepsilon}\\
     &=\frac{1}{2\varepsilon}\frac{\as ^2}{(2\pi)^2} C_FC_A\bigg\{\frac{1+x^2}{1-x}\Big[\frac{\pi^2}{3}-\frac{3}{2}\ln x-6\ln(1-x)-2\ln x\ln(1-x)+\frac{1}{2}\ln^2x-4\ln^2(1-x)+2\mathrm{Li}_2(1-x)\\
     &+6\ln\alpha+2\ln\alpha\ln(x(1-x))+2\ln^2\alpha\Big]+\frac{3}{2}-\frac{x}{2}+(3+x)\ln x+4x\ln(1-x)+3(1-x)\ln\alpha\bigg\}.
\end{aligned}
\end{equation}

The contribution from diagram A7 is
\begin{equation}
\begin{aligned}
    &f^-_\text{A7}\Big|_{1/\varepsilon}=2\as ^2C_FC_A(1-x)\int^{1-\alpha}_{1-x+\alpha}\mathrm{d}\xi(1-\xi)
    \left(\bar{\mu}^{4-D}\right)^2\int \frac{\mathrm{d}^{D-2}\mathbf{k}}{(2\pi)^{D-2}}\frac{1}{\left[\mathbf{k}^2+x(1-x)\mu^2\right]^2}\\
     &\times\int\frac{\mathrm{d}^{D-2}\mathbf{l}}{(2\pi)^{D-2}}\frac{[\bar{u}_{h}(p)\slashed\varepsilon_{\lambda}(q)u_{h'}(k)][\bar{u}_{h'}(k)\slashed\varepsilon_{\sigma'}(l')u_{s}(l'')][\bar{u}_{s}(l'')\slashed\varepsilon^*_{\sigma}(l)u_{h}(p)]\Gamma_{\sigma;\lambda,\sigma'}(\vec{l};\vec{q},\vec{l\,'})}{\mathbf{l}^2\left[\left(\mathbf{l}+\frac{1-\xi}{x}\mathbf{k}\right)^2-\frac{(1-\xi)(1-x-\xi)}{x^2(1-x)}\left(\mathbf{k}^2+x(1-x)\mu^2\right)\right]}\Big|_{1/\varepsilon}\\
     &=\frac{1}{2\varepsilon}\frac{\as ^2}{(2\pi)^2} C_FC_A\bigg\{\frac{1+x^2}{1-x}\Big[-\frac{1}{2}+\frac{\pi^2}{3}+\ln x+\frac{7}{2}\ln(1-x)+\ln x\ln(1-x)-\ln^2x+2\ln^2(1-x)-\mathrm{Li}_2(1-x)\\
     &+2\ln\alpha+\ln^2\alpha\Big]+\frac{1}{2}-x-4\ln x-2(1+x)\ln(1-x)+(1-x)\ln\alpha\bigg\}.
\end{aligned}
\end{equation}

Next, we consider the contributions from real emission diagrams contributing to the quark-quark splitting function, diagrams B1--B12. Diagram B1 reads
\begin{equation}
\begin{aligned}
    &f^-_\text{B1}\Big|_{1/\varepsilon}=\as ^2 C_FN_fT_F\int_0^{1-x}\mathrm{d}\xi\frac{x\xi(1-x-\xi)}{(1-x)^2}
    \left(\bar{\mu}^{4-D}\right)^2\int \frac{\mathrm{d}^{D-2}\mathbf{k}}{(2\pi)^{D-2}}\frac{1}{\left[\mathbf{k}^2+x(1-x)\mu^2\right]^2}\\
     &\times\int\frac{\mathrm{d}^{D-2}\mathbf{l}}{(2\pi)^{D-2}}\frac{[\bar{u}_{h}(p)\slashed\varepsilon_{\lambda}(q)u_{h'}(k)][\bar{v}_{s'}(l')\slashed\varepsilon_{\lambda}^*(q)u_{s}(l)][\bar{u}_{s}(l)\slashed\varepsilon_{\lambda'}(q)v_{s'}(l')][\bar{u}_{h'}(k)\slashed\varepsilon_{\lambda'}^*(q)u_h(p)]}{\left[\left(\mathbf{l}+\frac{\xi}{1-x}\mathbf{k}\right)^2+\frac{\xi(1-x-\xi)}{x(1-x)^2}\left(\mathbf{k}^2+x(1-x)\mu^2\right)\right]^2}\Big|_{1/\varepsilon}\\
     &=\frac{1}{2\varepsilon}\frac{\as ^2}{(2\pi)^2}C_FN_fT_F\bigg\{\frac{1+x^2}{1-x}\left[-\frac{8}{9}-\frac{2}{3}\ln x-\frac{4}{3}\ln (1-x)\right]-\frac{2}{3}(1-x)\bigg\}.
\end{aligned}
\end{equation}

Diagram B2 gives
\begin{equation}
\begin{aligned}
    &f^-_\text{B2}\Big|_{1/\varepsilon}=2\as ^2 C_FC_A\int_\alpha^{1-x-\alpha}\mathrm{d}\xi\frac{x\xi(1-x-\xi)}{(1-x)^2}
    \left(\bar{\mu}^{4-D}\right)^2\int \frac{\mathrm{d}^{D-2}\mathbf{k}}{(2\pi)^{D-2}}\frac{1}{\left[\mathbf{k}^2+x(1-x)\mu^2\right]^2}\\
     &\times\int\frac{\mathrm{d}^{D-2}\mathbf{l}}{(2\pi)^{D-2}}\frac{[\bar{u}_{h}(p)\slashed\varepsilon_{\lambda}(q)u_{h'}(k)][\bar{u}_{h'}(k)\slashed\varepsilon_{\lambda'}^*(q)u_h(p)]\Gamma_{\lambda';\sigma,\sigma'}(\vec{q};\vec{l},\vec{l}\,')\Gamma^\dagger_{\lambda;\sigma,\sigma'}(\vec{q};\vec{l},\vec{l}\,')}{\left[\left(\mathbf{l}+\frac{\xi}{1-x}\mathbf{k}\right)^2+\frac{\xi(1-x-\xi)}{x(1-x)^2}\left(\mathbf{k}^2+x(1-x)\mu^2\right)\right]^2}\Big|_{1/\varepsilon}\\
     &=\frac{1}{2\varepsilon}\frac{\as ^2}{(2\pi)^2}C_FC_A\bigg\{\frac{1+x^2}{1-x}\bigg[\frac{16}{9}+\frac{\pi^2}{3}+\frac{11}{6}\ln(x)-\frac{7}{3}\ln(1-x)-2\ln(x)\ln(1-x)-3\ln^2(1-x)\\
     &+6\ln\alpha+2\ln(\alpha)\ln(x(1-x))+\ln^2(\alpha) \bigg]+(1-x)\left[\frac{11}{6}+2\ln\left(\frac{\alpha}{1-x}\right)\right]\bigg\}.
\end{aligned}
\end{equation}

The contribution from the ladder diagram B3 is
\begin{equation}
\begin{aligned}
    &f^-_\text{B3}\Big|_{1/\varepsilon}=\as ^2 C_F^2\int_\alpha^{1-x-\alpha}\mathrm{d}\xi\frac{x\xi(1-x-\xi)}{(1-\xi)^2}
    \left(\bar{\mu}^{4-D}\right)^2\int \frac{\mathrm{d}^{D-2}\mathbf{l}}{(2\pi)^{D-2}}\frac{1}{\left[\mathbf{l}^2+\xi(1-\xi)\mu^2\right]^2}\\
     &\times\int\frac{\mathrm{d}^{D-2}\mathbf{l}}{(2\pi)^{D-2}}\frac{[\bar{u}_{h}(p)\slashed\varepsilon_{\lambda}(l)u_{s'}(l'')][\bar{u}_{s'}(l'')\slashed\varepsilon_{\lambda'}(l')u_{h'}(k)][\bar{u}_{h'}(k)\slashed\varepsilon^*_{\lambda'}(l')u_{s}(l'')][\bar{u}_{s}(l'')\slashed\varepsilon_{\lambda}^*(l)u_h(p)]}{\left[\left(\mathbf{k}+\frac{x}{1-\xi}\mathbf{l}\right)^2+\frac{x(1-x-\xi)}{\xi^2(1-\xi)}\left(\mathbf{l}^2+\xi(1-\xi)\mu^2\right)\right]^2}\Big|_{1/\varepsilon}\\
     &=\frac{1}{2\varepsilon}\frac{\as ^2}{(2\pi)^2}C_F^2\bigg\{\frac{1+x^2}{1-x}\Big[\frac{2\pi^2}{3}-12\ln(1-x)-4\ln x\ln(1-x)-6\ln^2(1-x)+12\ln\alpha+4\ln\alpha\ln(x(1-x))+2\ln^2\alpha\Big]\\
     &-4x\ln x+(1-x)\left[8+4\ln\alpha\right]+(1+x)\left[-2\ln x\ln(1-x)-\frac{3}{2}\ln^2x-2\mathrm{Li}_2(1-x)\right]\bigg\}.
\end{aligned}
\end{equation}
Note that this expression still contains the collinear subdivergence that is cancelled by the counterterm given in Eq.~\eqref{eq: matching counterterm}.

The calculation of diagram B4 is already presented in detail in section \ref{Section: Detailed example}. The result is Eq.~\eqref{Eq: diagram B4 result}.

The contribution from diagram B5 is
\begin{equation}
\begin{aligned}
    &f^-_\text{B5}\Big|_{1/\varepsilon}=-2\as ^2C_F\left(C_F-\frac{C_A}{2}\right)\int_{\alpha}^{1-x-\alpha}\mathrm{d}\xi\frac{\xi^2(1-\xi)}{1-x}
    \left(\bar{\mu}^{4-D}\right)^2\int \frac{\mathrm{d}^{D-2}\mathbf{k}}{(2\pi)^{D-2}}\frac{1}{\left[\mathbf{k}^2+x(1-x)\mu^2\right]}\\
     &\times\int\frac{\mathrm{d}^{D-2}\mathbf{l}}{(2\pi)^{D-2}}\frac{[\bar{u}_{h}(p)\slashed\varepsilon_{\lambda'}(l''')u_{s}(l)][\bar{u}_{s}(l)\slashed\varepsilon_{\lambda}(l'')v_{s'}(l')][\bar{v}_{s'}(l')\slashed\varepsilon^*_{\lambda'}(l''')u_{h'}(k)][\bar{u}_{h'}(k)\slashed\varepsilon_{\lambda}^*(l'')u_h(p)]}{\mathbf{l}^2\left[\mathbf{l}^2+\frac{x\xi}{1-x-\xi}\left(\mathbf{l}+\frac{1-\xi}{x}\mathbf{k}\right)^2\right]\left[\left(\mathbf{l}+\frac{\xi}{1-x}\mathbf{k}\right)^2+\frac{\xi(1-x-\xi)}{x(1-x)^2}\mathbf{k}^2\right]}\Big|_{1/\varepsilon}\\
     &=\frac{1}{2\varepsilon}\frac{\as ^2}{(2\pi)^2} C_F\left(C_F-\frac{C_A}{2}\right)\bigg\{\frac{1+x^2}{1-x}\Big[-8+2\mathrm{Li}_2(1-x)\Big]+1-\frac{2+6x+3x^2+5x^3}{(1-x)^2}\ln x\bigg\}.
\end{aligned}
\end{equation}

The contribution from diagram B6 is
\begin{equation}
\begin{aligned}
    &f^-_\text{B6}\Big|_{1/\varepsilon}=-\as ^2C_FC_A\int_{\alpha}^{1-x-\alpha}\mathrm{d}\xi\frac{\xi^2(1-\xi)}{1-x}
    \left(\bar{\mu}^{4-D}\right)^2\int \frac{\mathrm{d}^{D-2}\mathbf{k}}{(2\pi)^{D-2}}\frac{1}{\left[\mathbf{k}^2+x(1-x)\mu^2\right]}\\
     &\times\int\frac{\mathrm{d}^{D-2}\mathbf{l}}{(2\pi)^{D-2}}\frac{[\bar{u}_{h}(p)\slashed\varepsilon_{\lambda}(q)u_{h'}(k)][\bar{u}_{h'}(k)\slashed\varepsilon^*_{\sigma'}(l')u_{s}(l'')][\bar{u}_{s}(l'')\slashed\varepsilon^*_{\sigma}(l)u_{h}(p)]\Gamma^\dagger_{\lambda;\sigma,\sigma'}(\vec{q};\vec{l},\vec{l\,'})}{\mathbf{l}^2\left[\mathbf{l}^2+\frac{x\xi}{1-x-\xi}\left(\mathbf{l}+\frac{1-\xi}{x}\mathbf{k}\right)^2\right]\left[\left(\mathbf{l}+\frac{\xi}{1-x}\mathbf{k}\right)^2+\frac{\xi(1-x-\xi)}{x(1-x)^2}\mathbf{k}^2\right]}\Big|_{1/\varepsilon}\\
     &=\frac{1}{2\varepsilon}\frac{\as ^2}{(2\pi)^2} C_FC_A\bigg\{\frac{1+x^2}{1-x}\Big[-1-\frac{\pi^2}{6}+4\ln(1-x)-2\ln x\ln(1-x)-\ln^2x-\ln^2(1-x)-\mathrm{Li}_2(1-x)\\
     &-4\ln\alpha+2\ln\alpha\ln(x(1-x))-\ln^2\alpha\Big]-\frac{1}{2}+\frac{2+3x^2-x^3}{2(1-x)^2}\ln x\bigg\}.
\end{aligned}
\end{equation}

The remaining diagrams in this category are ones with two instantaneous interactions. The transverse momentum integrals in these diagrams are considerably easier, as they only include two energy denominators and no dependence on the loop momenta in the numerator. The result of diagram B7 is
\begin{equation}
\begin{aligned}
    &f^-_\text{B7}\Big|_{1/\varepsilon}=\as ^2 C_FN_fT_F\int_0^{1-x}\mathrm{d}\xi\frac{\xi(1-x-\xi)}{(p^+)^4x(1-x)^6}
    \\
     &\times\left(\bar{\mu}^{4-D}\right)^2\int \frac{\mathrm{d}^{D-2}\mathbf{k}}{(2\pi)^{D-2}}\int\frac{\mathrm{d}^{D-2}\mathbf{l}}{(2\pi)^{D-2}}\frac{[\bar{u}_{h}(p)\gamma^+u_{h'}(k)][\bar{v}_{s'}(l')\gamma^+u_{s}(l)][\bar{u}_{s}(l)\gamma^+v_{s'}(l')][\bar{u}_{h'}(k)\gamma^+u_h(p)]}{\left[\left(\mathbf{l}+\frac{\xi}{1-x}\mathbf{k}\right)^2+\frac{\xi(1-x-\xi)}{x(1-x)^2}\left(\mathbf{k}^2+x(1-x)\mu^2\right)\right]^2}\Big|_{1/\varepsilon}\\
     &=\frac{1}{2\varepsilon}\frac{\as ^2}{(2\pi)^2}C_FN_fT_F\bigg\{\frac{4}{3}\frac{x}{1-x}\bigg\}.
\end{aligned}
\end{equation}

The contribution from diagram B8 is
\begin{equation}
\begin{aligned}
    &f^-_\text{B8}\Big|_{1/\varepsilon}=\frac{1}{2}\as ^2 C_FC_A\int_0^{1-x}\mathrm{d}\xi\frac{\xi(1-x-\xi)(1-x-2\xi)^2}{(p^+)^2x(1-x)^6}
    \\
     &\times\left(\bar{\mu}^{4-D}\right)^2\int \frac{\mathrm{d}^{D-2}\mathbf{k}}{(2\pi)^{D-2}}\int\frac{\mathrm{d}^{D-2}\mathbf{l}}{(2\pi)^{D-2}}\frac{[\bar{u}_{h}(p)\gamma^+u_{h'}(k)][\bar{u}_{h'}(k)\gamma^+u_h(p)]\mathbf{\varepsilon}^*_{\lambda}\cdot\mathbf{\varepsilon}^*_{\lambda'}\mathbf{\varepsilon}_{\lambda}\cdot\mathbf{\varepsilon}_{\lambda'}}{\left[\left(\mathbf{l}+\frac{\xi}{1-x}\mathbf{k}\right)^2+\frac{\xi(1-x-\xi)}{x(1-x)^2}\left(\mathbf{k}^2+x(1-x)\mu^2\right)\right]^2}\Big|_{1/\varepsilon}\\
     &=\frac{1}{2\varepsilon}\frac{\as ^2}{(2\pi)^2}C_FC_A\bigg\{\frac{1}{3}\frac{x}{1-x}\bigg\}.
\end{aligned}
\end{equation}

The contribution from diagram B9 is
\begin{equation}
\begin{aligned}
    &f^-_\text{B9}\Big|_{1/\varepsilon}=\as ^2 C_F^2\int_0^{1-x}\mathrm{d}\xi\frac{\xi(1-x-\xi)}{(p^+)^2x(1-x)^2(1-\xi)^2}
    \\
     &\times\left(\bar{\mu}^{4-D}\right)^2\int \frac{\mathrm{d}^{D-2}\mathbf{k}}{(2\pi)^{D-2}}\int\frac{\mathrm{d}^{D-2}\mathbf{l}}{(2\pi)^{D-2}}\frac{[\bar{u}_{h}(p)\slashed \varepsilon_{\lambda}(l)\gamma^+\slashed\varepsilon_{\lambda'}(l') u_{h'}(k)][\bar{u}_{h'}(k)\slashed \varepsilon^*_{\lambda'}(l')\gamma^+\slashed\varepsilon^*_{\lambda}(l) u_{h}(p)]}{\left[\left(\mathbf{l}+\frac{1-\xi}{x}\mathbf{k}\right)^2+\frac{\xi(1-x-\xi)}{x(1-x)^2}\left(\mathbf{k}^2+x(1-x)\mu^2\right)\right]^2}\Big|_{1/\varepsilon}\\
     &=\frac{1}{2\varepsilon}\frac{\as ^2}{(2\pi)^2}C_F^2(1-x).
\end{aligned}
\end{equation}

Diagram B10 vanishes upon evaluating the numerator algebra, due to the diagram violating helicity conservation. Thus,
\begin{equation}
    f^-_\text{B10}\sim[\bar{u}_{h}(p)\slashed \varepsilon_{\lambda'}(l')\gamma^+\slashed\varepsilon_{\lambda}(l) u_{h'}(k)][\bar{u}_{h'}(k)\slashed \varepsilon^*_{\lambda'}(l')\gamma^+\slashed\varepsilon^*_{\lambda}(l) u_{h}(p)]=0.
\end{equation}
Note that the difference here to the numerator of diagram B9 are the gluon labels, which result in the numerator vanishing.

The contribution from diagram B11 is
\begin{equation}
\begin{aligned}
    &f^-_\text{B11}\Big|_{1/\varepsilon}=-2\as ^2 C_F\left(C_F-\frac{C_A}{2}\right)\int_0^{1-x}\mathrm{d}\xi\frac{\xi}{(p^+)^2x(1-x)^3(1-\xi)}
    \\
     &\times\left(\bar{\mu}^{4-D}\right)^2\int \frac{\mathrm{d}^{D-2}\mathbf{k}}{(2\pi)^{D-2}}\int\frac{\mathrm{d}^{D-2}\mathbf{l}}{(2\pi)^{D-2}}\frac{[\bar{u}_{h}(p)\gamma^+u_{s}(l)][\bar{u}_{s}(l)\gamma^+v_{s'}(l')][\bar{v}_{s'}(l')\gamma^+u_{h'}(k)][\bar{u}_{h'}(k)\gamma^+u_h(p)]}{\left[\mathbf{l}^2+\frac{x\xi}{1-x-\xi}\left(\mathbf{l}+\frac{1-\xi}{x}\mathbf{k}\right)^2\right]\left[\left(\mathbf{l}+\frac{\xi}{1-x}\mathbf{k}\right)^2+\frac{\xi(1-x-\xi)}{x(1-x)^2}\left(\mathbf{k}^2+x(1-x)\mu^2\right)\right]}\Big|_{1/\varepsilon}\\
     &=\frac{1}{2\varepsilon}\frac{\as ^2}{(2\pi)^2}C_F\left(C_F-\frac{C_A}{2}\right)\left\{16\frac{x}{1-x}+8\frac{x(1+x)}{(1-x)^2}\ln x\right\}.
\end{aligned}
\end{equation}

The contribution from diagram B12 is
\begin{equation}
\begin{aligned}
    &f^-_\text{B12}\Big|_{1/\varepsilon}=\frac{1}{2}\as ^2 C_FC_A\int_0^{1-x}\mathrm{d}\xi\frac{\xi(1-x-2\xi)}{(p^+)^2x(1-x)^3}
    \\
     &\times\left(\bar{\mu}^{4-D}\right)^2\int \frac{\mathrm{d}^{D-2}\mathbf{k}}{(2\pi)^{D-2}}\int\frac{\mathrm{d}^{D-2}\mathbf{l}}{(2\pi)^{D-2}}\frac{[\bar{u}_{h}(p)\slashed \varepsilon_{\lambda}(l)\gamma^+\slashed\varepsilon_{\lambda'}(l') u_{h'}(k)][\bar{u}_{h'}(k)\gamma^+u_h(p)]\mathbf{\varepsilon}^*_{\lambda}\cdot\mathbf{\varepsilon}^*_{\lambda'}}{\left[\mathbf{l}^2+\frac{x\xi}{1-x-\xi}\left(\mathbf{l}+\frac{1-\xi}{x}\mathbf{k}\right)^2\right]\left[\left(\mathbf{l}+\frac{\xi}{1-x}\mathbf{k}\right)^2+\frac{\xi(1-x-\xi)}{x(1-x)^2}\left(\mathbf{k}^2+x(1-x)\mu^2\right)\right]}\Big|_{1/\varepsilon}\\
     &=\frac{1}{2\varepsilon}\frac{\as ^2}{(2\pi)^2}C_FC_A\left\{-2\frac{x}{1-x}-\frac{x(1+x)}{(1-x)^2}\ln x\right\}.
\end{aligned}
\end{equation}

Finally, we consider the contributions to the quark-antiquark splitting function, given by diagrams C1 and C2. The contribution from diagram C1 is
\begin{equation}
\begin{aligned}
    &f^-_\text{C1}\Big|_{1/\varepsilon}=\as ^2C_F\left(C_F-\frac{C_A}{2}\right)\int_{0}^{1-x}\mathrm{d}\xi\frac{(x+\xi)(1-x-\xi)^2}{1-\xi}
    \left(\bar{\mu}^{4-D}\right)^2\int \frac{\mathrm{d}^{D-2}\mathbf{l}}{(2\pi)^{D-2}}\frac{1}{\left[\mathbf{l}^2+\xi(1-\xi)\mu^2\right]}\\
     &\times\int\frac{\mathrm{d}^{D-2}\mathbf{l}}{(2\pi)^{D-2}}\frac{[\bar{u}_{h}(p)\slashed\varepsilon_{\lambda'}(l''')u_{s'}(l')][\bar{u}_{s'}(l')\slashed\varepsilon_{\lambda}(l'')v_{h'}(k)][\bar{v}_{h'}(k)\slashed\varepsilon^*_{\lambda'}(l''')u_{s}(l)][\bar{u}_{s}(l)\slashed\varepsilon_{\lambda}^*(l'')u_h(p)]}{\left(\mathbf{k}+\mathbf{l}\right)^2\left[\left(\mathbf{k}+\mathbf{l}\right)^2+\frac{\xi(1-x-\xi)}{x}\left(\mathbf{k}-\frac{x}{\xi}\mathbf{l}\right)^2\right]\left[\left(\mathbf{k}+\frac{x}{1-\xi}\mathbf{l}\right)^2+\frac{x(1-x-\xi)}{\xi(1-\xi)^2}\mathbf{l}^2\right]}\Big|_{1/\varepsilon}\\
     &=\frac{1}{2\varepsilon}\frac{\as ^2}{(2\pi)^2}C_F\left(C_F-\frac{C_A}{2}\right)\left\{\left[\frac{\pi^2}{3}+4\ln x\ln(1+x)-\ln^2x+4\mathrm{Li}_2(-x)\right]\frac{1+x^2}{1+x}-4\frac{(1-x^4)}{(1+x)^3}-2\frac{1+6x^2+x^4}{(1+x)^3}\ln x\right\}.
\end{aligned}
\end{equation}

The contribution from diagram C2 is
\begin{equation}
\begin{aligned}
    &f^-_\text{C2}\Big|_{1/\varepsilon}=\as ^2C_F\left(C_F-\frac{C_A}{2}\right)\int_{0}^{1-x}\mathrm{d}\xi\frac{1-x-\xi}{(p^+)^2\xi(1-\xi)^3(x+\xi)}
    \\
     &\times\left(\bar{\mu}^{4-D}\right)^2\int \frac{\mathrm{d}^{D-2}\mathbf{l}}{(2\pi)^{D-2}}\int\frac{\mathrm{d}^{D-2}\mathbf{l}}{(2\pi)^{D-2}}\frac{[\bar{u}_{h}(p)\gamma^+u_{s'}(l')][\bar{u}_{s'}(l')\gamma^+v_{h'}(k)][\bar{v}_{h'}(k)\gamma^+u_{s}(l)][\bar{u}_{s}(l)\gamma^+u_h(p)]}{\left[\left(\mathbf{k}+\mathbf{l}\right)^2+\frac{\xi(1-x-\xi)}{x}\left(\mathbf{k}-\frac{x}{\xi}\mathbf{l}\right)^2\right]\left[\left(\mathbf{k}+\frac{x}{1-\xi}\mathbf{l}\right)^2+\frac{x(1-x-\xi)}{\xi(1-\xi)^2}\mathbf{l}^2\right]}\Big|_{1/\varepsilon}\\
     &=\frac{1}{2\varepsilon}\frac{\as ^2}{(2\pi)^2} C_F\left(C_F-\frac{C_A}{2}\right)\bigg\{-8\frac{x(1-x)}{(1+x)^2}-8\frac{x(1+x^2)}{(1+x)^3}\ln x\bigg\}.
\end{aligned}
\end{equation}

\bibliographystyle{JHEP-2modlong}
\bibliography{references}

\end{document}